\documentclass[aps,pre,preprint,superscriptaddress]{revtex4-1}

\usepackage{amsmath,amsthm,amssymb,color,xcolor,mathtools,amsfonts,bm,chemfig,hhline,subfig}
\usepackage[ pdfborderstyle={/S/U/W 1}]{hyperref}
\definecolor{MyBlue}{HTML}{210cac}
\definecolor{MyCiteColor}{HTML}{0099FF}
\definecolor{MyRed}{HTML}{3E186A}
\hypersetup{%
  pdfborderstyle={/S/U/W 1},
  colorlinks=true,
  linkcolor=MyBlue,
  citecolor=MyCiteColor,
  urlcolor=MyRed,
  anchorcolor=MyBlue,
  linkbordercolor=MyRed,
}
\usepackage[shortlabels]{enumitem}
\usepackage{graphicx}
\usepackage{makecell}
\usepackage{tikz}
\usepackage[normalem]{ulem}

\usepackage{caption}
\newcommand{\RR}{\mathbb{R}}

\newcommand \bk{\color{black}}

\usepackage[normalem]{ulem}
\graphicspath{{Figures/}}

\begin{document}
\title{Coloured Noise from Stochastic Inflows in Reaction-Diffusion Systems}

\author{Michael F Adamer}
\email{adamer@maths.ox.ac.uk}
\affiliation{Wolfson Centre for Mathematical Biology, Mathematical Institute, University of Oxford}

\author{Heather A Harrington}
\affiliation{Wolfson Centre for Mathematical Biology, Mathematical Institute, University of Oxford}

\author{Eamonn A Gaffney}
\affiliation{Wolfson Centre for Mathematical Biology, Mathematical Institute, University of Oxford}

\author{Thomas E Woolley}
\affiliation{Cardiff School of Mathematics, Cardiff University}

\date{\today}

\begin{abstract}
In this paper we present a framework for investigating coloured noise in reaction-diffusion systems. We start by considering a deterministic reaction-diffusion equation and show how external  forcing can cause temporally correlated or coloured noise. Here, the main source of external noise is considered to be fluctuations in the parameter values representing the inflow of particles to the system.
First, we determine which reaction systems, driven by extrinsic noise, can admit only one steady state, so that effects, such as stochastic switching, are precluded from our analysis. To analyse the steady state behaviour of reaction systems, even if the parameter values are changing, necessitates a parameter-free approach, which has been central to algebraic analysis in chemical reaction network theory.
To identify suitable models we use tools from real algebraic geometry that link the network structure to its dynamical properties. We then make a connection to internal noise models and show how power spectral methods can be used to predict stochastically driven patterns in systems with coloured noise. In simple cases we show that the power spectrum of the coloured noise process and the power spectrum of the reaction-diffusion system modelled with white noise multiply to give the power spectrum of the coloured noise reaction-diffusion system.
\end{abstract}
\maketitle

\section{Introduction}

One of the central challenges in mathematical biology is understanding mechanisms involved in development processes. Within the context of developmental biology, the emergence of large scale spatial structure, has often been theoretically explored through a common framework of deterministic partial differential equations defining reaction-diffusion systems
\cite{Murray2008,Hochberg2003,Turing37}.
While current frameworks may explain  a variety of phenomena in development, they can also suffer from over-simplification \cite{McKane2014}, with the additional caveat that finding theoretical models both consistent with mechanism-based knowledge and capable of predicting observed patterns is a highly complex task, suffering from both model and parameter fine tuning \cite{Butler2011,Xavier2013}.

Many models that describe pattern formation assume parameters are constant; however, this deterministic assumption is not suitable for certain conditions. Some systems are better suited towards a stochastic approach. When a system is coupled to external and stochastic drivers, then the parameter values can change. The stochastic driving is often represented by stochastic parameters, that is a parameter which is drawn from a certain distribution at each time step or spatial point, and referred to as {\it extrinsic noise} below.
This contrasts with most previous work on stochastic pattern formation, referred to as {\em intrinsic noise}, which assumes low copy number and it does not assume external drivers as the source of noise \cite{Schumacher2013,McKane2005,McKane2014,Biancalani2011,woolley2012effects,woolley2011influence}.

The structure of intrinsic noise is often taken to be be highly constrained and in particular uncorrelated in time, leading to white noise representations. For instance, if the noise is due to low copy number induced dynamics, Gaussian white noise forcing emerges from the chemical Langevin equation approximation  to the chemical master equation \cite{Gillespie2000}. Even with such constraints
there is a rich diversity of noise-induced phenomena, such as spatio-temporal pattern formation \cite{Schumacher2013,woolley2012effects}, stochastic oscillations \cite{McKane2005}, metastability \cite{McKane2014}, waves \cite{Biancalani2011} or enhanced oscillation amplitude \cite{Dauxois2009}.

However, when the source of the noise is {\em extrinsic}, other forms of noise are permissible. In particular, the defining special properties of white noise
may in general be relaxed and hence stochastic forcing can emerge with  non-trivial temporal correlations, often termed {\em coloured noise}. Our objective is to show how extrinsic noise influences
spatio-temporal reaction-diffusion patterns, in particular by  developing  power spectral methods to analyse the impact of coloured noise.

To proceed, we first note deterministic pattern formation reaction diffusion systems in biological applications with $n$ interacting biochemical species take the form \cite{Turing37,Murray2008}
\begin{equation}
  \frac{\partial \bm{x}}{\partial t} = \bm{D}\frac{\partial^2 \bm{x}}{\partial s^2} + \bm{f}(\bm{x})
  \label{GenericRD}
\end{equation}
in one spatial dimension, where $\bm{x} \in \RR^n_{\geq 0}$ is a vector of $n$ chemical concentrations. The models we consider require the term $\bm{f}(\bm{x}) = (f_1(\bm{x}),\dots,f_n(\bm{x}))^T$ to be a vector of rational functions. The rational functions describe the underlying reaction network between the species
and $\bm{D}(\partial^2 \bm{x}/\partial s^2)$ with $\bm{D} = \text{diag}(D_1,\dots,D_n)$ describes the diffusion of $\bm{x}$. Let there be $M$ chemical reactions in the spatially homogeneous system whose dynamics are described by $\partial \bm{x}/\partial t = \bm{f}(\bm{x})$. Each reaction is parametrised by a reaction rate $k_i > 0$ and therefore we have a vector of reaction rates $\bm{k}=(k_1,\dots,k_M) \in \RR^M_{> 0}$. Throughout this paper we restrict our analysis to {\em mass action kinetics} \cite{Horn1972}, rendering $\bm{f}(\bm{x})$ a vector of polynomials and we use homogeneous Neumann boundary conditions
\begin{equation}
  \frac{\partial \bm{x}}{\partial s}\biggr\rvert_{s=0,L} = 0\:,
\end{equation}
where $L$ is the domain length. Note that, although our work is only applied to systems of one spatial dimension, the theory can easily be extended to an arbitrary number of spatial dimensions.

Counter-intuitively, it has been shown that under certain conditions \cite{Murray2008} diffusion can drive an otherwise spatially uniform stable state to instability. Such unstable systems, which can form stable patterns, such as stripes or spots, are called {\em Turing Systems} \cite{Turing37,woolley2017turing}.
To avoid the complications of bistable dynamics, and later stochastic analogues such as switching between steady states, we impose the constraint that the spatially homogeneous solution has a unique stable steady state, $\bm{x}^*$, such that $\bm{f}(\bm{x}^*) = 0$ \cite{Murray2008}. To find models which have this property we use techniques from real algebraic geometry \cite{Muller2016} which provides simple tools, which ensure there exists only a single steady state in a model. Since Turing's work in 1952 many biological patterning systems have been suggested to be Turing systems \cite{Castets1990,Cartwright2002,Kondo2010}.

Fundamentally, the application of a set of partial differential equation (PDE) models for  a  biochemical reaction system assumes that the species of interest are in high enough concentration to allow continuum modelling. By contrast, when the number of particles in the biological system is low, intrinsic stochasticity of the system must be included in the model  \cite{Schumacher2013,woolley2012effects,McKane2005,McKane2014,Biancalani2011}, which typically yields studies that investigate  the impact of white noise.

However, as mentioned, stochasticity in biological systems can also arise from extrinsic noise and hence temporal variations in parameter values \cite{Picco2017}. Thus, instead  we generalise the deterministic system to include {\em stochastic} parameter values. Specifically, we focus on the effect of stochastic fluctuations in the constant term, the ``inflow'' term, of the chemical reaction network $\bm{f}(\bm{x})$. As the extrinsic noise can arise from a vast number of different mechanisms, such as varying experimental conditions \cite{Lenive2016}, it is largely free of microscopic constraints, especially the absence of temporal correlation.
However, the impact of correlated external noise has, to the authors' knowledge,  received little theoretical modelling attention. Thus we proceed to develop a framework to study external coloured noise forcing of the above deterministic system for pattern formation in biologically motivated scenarios, analysing how temporal correlation in noise, as described by colour, impacts self-organisation properties.

The paper is organised as follows.
In section \ref{Theory} we introduce required notions of chemical reaction network theory to select a class of models relevant to our framework, i.e. those whose number of steady states is unaltered by changes in parameter values, temporal or otherwise. We then introduce stochastic inflow parameters and describe their connection coloured noise. In section \ref{Computations} we highlight the impact of noise colouring on the spatio-temporal patterns formed by example of the Schnakenberg system. In particular, we discuss noise arising from stochastic subnetworks and varying experimental conditions. Our numerical results are discussed in section \ref{Conclusion} where we summarise the distinguishable differences between the various noise colours.

\section{Theoretical Results}\label{Theory}

In this section we introduce the theoretical background of this paper.
In subsection \ref{CRNT} we introduce chemical reaction networks and use results from real algebraic geometry to select models where changing a parameter does not effect the uniqueness of the steady state of the chemical reaction network. Networks that satisfy a unique steady state are required for our framework.
In subsection \ref{RandomInflows} we define stochastic inflows and describe their potential sources. In subsection \ref{PowerSpectra} we first show how the generic law of mass action reaction-diffusion system can be approximated to highlight the mathematical equivalence of external and internal noise in our framework. Next, we introduce power spectra as an analytic tool to study spatial and temporal pattern formation and, finally, we calculate a general formula for power spectra which includes the effects coloured noise.

\subsection{Chemical Reaction Network Theory}\label{CRNT}

A central aim of chemical reaction network theory (CRNT) is to describe the properties of a chemical reaction network from its reaction graph alone \cite{Feinberg1987,Feinberg1988}. One such property is the capacity for multiple steady states. Define a chemical reaction network by the multi-set $\mathcal{N} = \{\mathcal{S}, \mathcal{C}, \mathcal{R}\}$, where $\mathcal{S}, \mathcal{C}, \mathcal{R}$ are defined below. We begin by embedding the network into $n$ dimensional Euclidean space $\RR^n$ by associating a basis vector $\bm{e}_i$ to each chemical species $X_i$ such that $X_1\to \bm{e}_1 = (1,0,0,\dots)^T$, $X_2\to \bm{e}_2 = (0,1,0,\dots)^T$ and so on. Let $\mathcal{S} = \{X_1,\dots,X_n\}$ be the set of all chemical species in the network, then a generic reaction can be expressed as
\begin{equation}
  \sum_{i=1}^n \alpha_i X_i \xrightarrow{k} \sum_{i=1}^n \beta_i X_i\; .
  \label{GenericReaction}
\end{equation}
The constants $\alpha_i$ and $\beta_i$ are {\em stoichiometric coefficients} which give information about how many molecules of $X_i$ are consumed and produced in each reaction. By letting $X_i\to \bm{e}_i$ we can formulate a {\em reaction vector} describing the consumption, or production, of a species $X_i$ in a reaction
\begin{equation}
  \bm{r} = \sum_{i=1}^n (\beta_i-\alpha_i)\bm{e}_i\; .
\end{equation}
If an entry of $\bm{r}$ is negative, then a species is consumed, whilst if an entry of $\bm{r}$ is positive, then a species is produced.

Denote the set of all reaction vectors in a network by $\mathcal{R} = \{\bm{r}_1,\dots,\bm{r}_M\}$. To complete the description of a chemical reaction network in terms of sets and their embedding into Euclidean space we introduce the notion of {\em complexes}. Complexes are linear combinations of species which appear on the left or right hand sides of reaction vectors. In equation \eqref{GenericReaction} the two complexes are $C_1 = \sum_{i=1}^n \alpha_i X_i$ and $C_2 = \sum_{i=1}^n \beta_i X_i$, where $C_1$ is the {\em reactant complex} and $C_2$ is the {\em product complex}.
Let the set of all complexes be $\mathcal{C} = \{C_1,\dots,C_l\}$. The reaction network, $\mathcal{N}$, is therefore a directed graph with vertex set $\mathcal{C}$ and a directed edge between vertices $C_i$ and $C_j$ if and only if $C_j-C_i \in \mathcal{R}$, with the same embedding, $X_i\to \bm{e}_i$.
Note that the description of the reaction network {\em does not} include any notion of rate constants $k$, however, many results in chemical reaction network theory include the reaction rates explicitly as a positive vector $\bm{k} = (k_1,\dots,k_m)^T\in\RR^M_{>0}$.

In this paper we study the influence of noise on reaction-diffusion systems that are able to produce patterns in a well-defined parameter region. Critically, we will see that coloured noise is able to have both a constructive influence outside of this previously defined parameter region (i.e. the noise stabilises patterns where we would not expect them deterministically), as well as a destructive influence  on patterns which normally would arise. In particular, we would like to exclude the intrinsically stochastic effect of {\em stochastic switching}, which occurs for a system that has multiple steady states in some parameter region. Stochastic switching describes the phenomenon that a chemical reaction network can jump from one steady state to another when subject to finite stochastic perturbations. To avoid this scenario we use a network structure based tool described in \cite{Muller2016} which a priori excludes multistationarity.

Take a chemical reaction network $\mathcal{N}$ and embed its complexes and reactions into $\mathbb{R}^n$. Then define the matrices $m_1 = (\bm{r}_1 \bm{r}_2 \cdots \bm{r}_M)$ for $r_i\in \mathcal{R}$ and $m_2^T = (\bm{b}_1 \bm{b}_2\cdots \bm{b}_r)$ where $\{\bm{b}_i\}$ is the set of reactant complex vectors of $\mathcal{N}$. Let $diag(k_1,\dots,k_M)$ be the diagonal matrix of reaction constants. Using the law of mass action the dynamics of the species concentrations can be described by
\begin{equation}
  \frac{d\bm{x}}{dt} = \bm{f}(\bm{x}) = m_1\:diag(k_1,\dots,k_m)\: \bm{x}^{m_2}
\end{equation}
where $\bm{x}^{m_2} = (x_1^{b_{11}}x_2^{b_{12}}\cdots x_n^{b_{1n}},\dots,x_1^{b_{M1}}x_2^{b_{M2}}\cdots x_n^{b_{Mn}})^T$. Note that the reactant complexes and the reactions are structural properties of the reaction graph.
Further, let $\bm{a} \in \mathbb{R}^n$ and define its sign vector $\sigma(\bm{a}) = \{-,0,+\}^n$ by applying the sign function to each component of $\bm{a}$. Therefore, the $i^\text{th}$ component of the sign vector $\sigma(\bm{a})_i = \text{sign}(a_i) \in \{+,0,-\}$.

The link from network structure to multistationarity is outlined in \cite[Theorem 1.4]{Muller2016} and it concerns the injectivity of the map  $\bm{f}: \bm{x}\mapsto d\bm{x}/dt$. If $\bm{f}(\bm{x})$ is injective then there exists a unique vector $\bm{x}^*\in \mathbb{R}^n$ such that $\bm{f}(\bm{x}^*) = 0$. In other words the network is monostationary. The conditions for an injective $\bm{f}(\bm{x})$ hold for all parameter values $\bm{k} = (k_1,\dots,k_M)$. A corollary of \cite[Theorem 1.4]{Muller2016} is that $\bm{f}(\bm{x})$ is injective if
\begin{equation}
  \text{ker}(m_2) = \{0\}\quad\quad \text{and}\quad\quad \sigma(\text{ker}(m_1))\cap\sigma(\text{im}(m_2)) = \{0\}\: .
  \label{Inject}
\end{equation}
Note, that this condition depends on the network structure, specifically, the spaces spanned by the kernel of the reaction matrix, $m_1$, and the image of the source complex matrix, $m_2$, but not on the parameter values.

\subsection{Stochastic Inflow in Chemical Reaction Networks}\label{RandomInflows}

We will now show how chemical reaction networks (CRNs) that satisfy injectivity (preclusion of multistationarity) can be applied to parameter stochasticity, in particular, to the case of stochastic inflows.

Much effort in the CRNT literature investigates the effect of inflows into the chemical system, especially when the network is changed to a ``fully open system'' in which every species has an inflow reaction \cite{Banerjee2014,Joshi2014,Felix2016,Banaji2018}. By contrast, here we change the nature of the model by generalising the inflow reaction rates to stochastic processes, rather than simply adding inflow processes as deterministic reactions. Sometimes these inflow reactions are the called``inputs'' in engineering applications and the concentrations are the ``outputs'' and, therefore the study of chemical reaction networks with varying inflows can be rephrased into a study of the ``input-output'' relation as has been the approach in previous work \cite{Frank2013}.

Consider a chemical reaction network $\mathcal{N}$. An {\em inflow} reaction for species $X_i$ is manifested in the reaction graph as
\begin{equation*}
  \emptyset \xrightarrow{k_{\text{in}}} X_i\;.
\end{equation*}
Usually, the work in CRNT assumes the reaction rate $k_{\text{in}}$ to be constant in time (and space). However, the zero complex, $\emptyset$, symbolises a process not included in the model such as the the production of $X_i$ by another network, which we call an ``auxiliary network'', or a mechanical addition of $X_i$ to an experimental setup. Both mechanisms are often subsumed into $\emptyset$ and usually approximated as constant or ``perfect'' inflow, however, they can exhibit intricate dynamics. We model the dynamics of ``non-perfect'' inflows by a stochastic process whose origin can be two-fold.
\begin{enumerate}
  \item[(a)] {\em Stochastic sub-systems}: We assume that the inflow into the (deterministic) reaction diffusion system is provided by the output of another chemical reaction network (with a unique fixed point). When the number of particles in the {\em auxiliary network} is large the system will be in a steady state and the inflow rate $k_\text{in}$ will simply be proportional to the steady state value of a species of the auxiliary system. However, when the particle number in the auxiliary system is low, stochastic fluctuations cannot be ignored and, while still proportional to the concentration of a species in the subsystem, the actual influx parameter $k_\text{in}$ will be a stochastic process $K_\text{in}(t)$.
  \item[(b)] {\em Experimental fluctuations}: In chemical engineering it is assumed that inflow of chemicals into a reactor is a perfectly deterministic process, however, due to mechanical or other experimental imperfections the inflow into a reaction can vary stochastically. This again renders the influx parameter $k_{\text{in}}$ into a stochastic process $K_\text{in}(t)$.
\end{enumerate}

\begin{figure}
  \includegraphics[width=.5\textwidth]{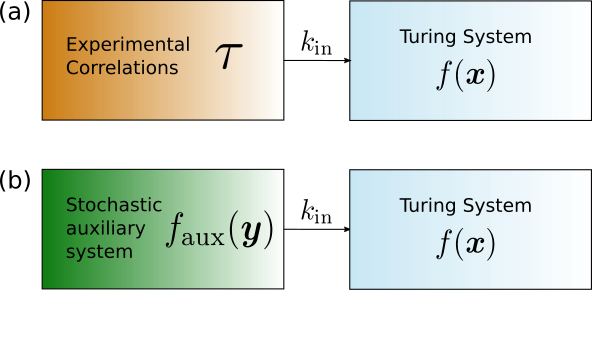}
  \caption{\label{Concepts} The two mechanisms contributing to stochastic inflows. The boxes on the left are usually treated as black boxes resulting in some constant inflow modelled by the parameter $k_\text{in}$. We differentiate between experimental fluctuations, symbolised by a correlation $\tau$ in (a) and auxiliary networks described by $f_\text{aux}(\bm{y})$ in case (b).}
\end{figure}

Whereas the two sources of parameter noise are conceptually different, their mathematical description is the same.
Consider a stochastic process $K(t)$ with an underlying distribution $\mathcal{D}(t)$. Therefore, at every time $t$ we have $K(t)\sim \mathcal{D}(t)$.
To be biologically relevant (i.e. to ensure all chemical concentrations are non-negative at every $t$) we require $\mathcal{D} = 0$ for all $K(t) < 0$.
To simplify the following mathematical analysis we approximate the distribution of $K(t)$ as a {\em truncated Gaussian}, since we would like to connect our analysis to the internal noise case.
Therefore, we can describe $K(t)$ as
\begin{align}
  K(0) &= k,\nonumber\\
  \Delta K &= K(t) - \mathbb{E}\left[K(t')\right] = \frac{1}{\sqrt{\Omega}} \sqrt{g(t,t')}\xi,\nonumber\\
  K(t) &= 0\qquad\text{if }\Delta K < -\mathbb{E}\left[K(t')\right],\nonumber\\
  \intertext{such that:}
  \mathbb{E}[K(t)] &= k\quad\forall t \geq 0,
\end{align}
where $\xi \sim N(0,1)$, $\Omega$ is positive constant and $g(t,t')$ is a positive function.
We assume that the standard deviation of the Gaussian perturbation to the mean parameter value is small such that $K(t)<0$ is a rare event.
Due to the negligible truncation, we approximate the moments of the truncated Gaussian as the moments of the full Gaussian such that $\langle K(t)-k\rangle = 0$ and $\langle (K(t)-k)(K(t')-k)\rangle = \Omega^{-1} g(t,t')$.
In the remainder of this paper we will show how various functions $g(t,t')$ can arise in mathematical modelling, especially when stochastic auxiliary networks are considered.

We can further connect the parameter $\sigma$ to physical quantities of the underlying chemical reaction network by again considering the sources of stochastic inflow. We assume that the origin of the stochastic inflow is stochastic auxiliary networks or experimental imperfections. Therefore, in the limit of either a perfect experiment or a deterministic auxiliary network the inflow should be deterministic and $\langle(K(t)-k)(K(t')-k)\rangle = 0$ implying $\Omega \to \infty$. For stochastic auxiliary networks the quantity $\Omega$ which parametrises the size of the stochastic fluctuations can be related to the {\em system size} \cite{Kampen2007}. Similarly to the system size expansion, we require $\Omega$ to be sufficiently large. Large $\Omega$ helps to maintain a positive solution of the stochastic partial differential equation, although, positivity cannot be guaranteed when the noise is correlated.

Next, consider a chemical reaction network $\mathcal{N}$ in which a subset of species has an inflow reaction. Denote the set of species with (stochastic) inflow by $\mathcal{S}_\text{in} \subseteq \mathcal{S}$. If there is more than one (stochastic) inflow $|\mathcal{S}_\text{in}|>1$ the stochastic process will be multi-dimensional Brownian motion. The description of the stochastic process outlined in this section still applies in this case, if we have $\bm{K}_\text{in}(t)$ as the vector of stochastic inflow such that
\begin{align}
  \mathbb{E}[\bm{K}_\text{in}(t)] &= \bm{k}_\text{in}\nonumber
  \intertext{and}
  \bm{K}_\text{in}(t)-\mathbb{E}\left[\bm{K}_\text{in}(t')\right]&= \bm{\eta}\nonumber
  \intertext{with}
  \langle\bm{\eta}\rangle &= 0\nonumber\\
  \langle\bm{\eta}\bm{\eta}^T\rangle &= BG(t,t') = \{B_{ij}G_{ij}(t,t')\}
  \label{Correlations}
  \end{align}
and $B$ representing the potential covariance between the stochastic inflows.

Since our assumptions made the inflow process additive and uncoupled to the species, the stochastic reaction network can be described by the system of  stochastic partial differential equations (SPDEs)
\begin{equation}
  \frac{\partial \bm{x}}{\partial t} = D\frac{\partial^2 \bm{x}}{\partial s^2} + \bm{f}(\bm{x}) + \frac{1}{\sqrt{\Omega}}\bm{\eta}\; ,
  \label{StochNet}
\end{equation}
where $\bm{\eta}$ is a vector of stochastic processes such that its support is $supp(\bm{\eta}) = \mathcal{S}_\text{in}$.

Note that in our limit randomising inflows leaves $\mathcal{N}$ as well as $\bm{f}(\bm{x})$ invariant, and hence, the steady state structure of the system; i.e. if we start out with an injective function $\bm{f}(\bm{x})$ then adding stochasticity will not change this injectivity.

\subsection{Power Spectra for Stochastic Inflows}\label{PowerSpectra}

To fully classify the patterns arising from the addition of stochastic inflows we calculate the power spectra of the patterns. Power spectra are an analytic tool showing which spatial and temporal frequencies are present in a pattern \cite{Adamer2017,Schumacher2013,Woolley2011a}. Peaks in power spectra give information about dominant frequencies and, hence, about oscillatory behaviour of a system in space and time.

First, we linearise equation \eqref{StochNet} about the fixed point $\bm{x} \sim \bm{x}^* + \overline{\delta \bm{x}}$. Where $\overline{\delta \bm{x}}$ represent small perturbations; these should decay in the deterministic limit as the system will converge to the steady state outside the Turing parameter regime. Inside the Turing regime the system is assumed to converge to a stable pattern rather than a stable state which is not considered in this paper. Hence, our analysis is valid only {\em outside} the Turing pattern regime. Substituting the linearisation ansatz into equation \eqref{StochNet} and keeping lowest order terms only we get
\begin{equation}
  \frac{\partial \overline{\delta \bm{x}}}{\partial t} = D\frac{\partial^2 \overline{\delta \bm{x}}}{\partial s^2} + J|_{\bm{x}=\bm{x}^*}\overline{\delta \bm{x}} + \frac{1}{\sqrt{\Omega}}\bm{\eta}\; ,
  \label{StochNetLin}
\end{equation}
where $J|_{\bm{x}=\bm{x}^*}$ is the Jacobian of $\bm{f}(\bm{x})$ evaluated at the fixed point $\bm{x}^*$, which for notational convenience we will denote as $J$.

We make one further assumption, namely the scaling of the perturbation $\overline{\delta \bm{x}}$ with the parameter $\Omega$ controlling the magnitude of the stochastic input. Let $\overline{\delta \bm{x}} = \Omega^\alpha \delta \bm{x}$. Then, if $\alpha < -1/2$ and in the limit of $\Omega \rightarrow \infty$ the leading order term is the stochastic process $\bm{\eta}$ only, similarly, if $\alpha > -1/2$, then $\bm{\eta}$ could be neglected to leading order in $\Omega$. Hence, we let $\alpha = 1/2$ and equation \eqref{StochNetLin} simplifies to
\begin{equation}
  \frac{\partial \delta \bm{x}}{\partial t} = D\frac{\partial^2 \delta \bm{x}}{\partial s^2} + J\delta \bm{x} + \bm{\eta}\; ,
  \label{StochNetLinFinal}
\end{equation}

Note, that equation \eqref{StochNetLinFinal} is mathematically equivalent to a chemical Langevin equation of compartmentalised diffusion in the limit of the compartment size, $\Delta_s$, going to zero \cite{Smith1985}. However, in our derivation the source of the noise is {\em external}. To emphasise the mathematical connection with internal noise we will in fact discretise equation \eqref{StochNetLinFinal} into a finite number of compartments, such that for an $n$ species network with $\mathcal{K}$ compartments we have
\begin{subequations}
\begin{align}
  \delta \bm{x} &= [\delta x_1,\dots,\delta x_\mathcal{K},\delta x_{\mathcal{K}+1},\dots, \delta x_{2\mathcal{K}},\dots,\delta x_{n\mathcal{K}}]^T\:,\\
  \bm{\eta} &= [\eta_1,\dots,\eta_\mathcal{K},\eta_{\mathcal{K}+1},\dots, \eta_{2\mathcal{K}},\dots,\eta_{n\mathcal{K}}]^T\:,\\
  s &\approx s_i = i\Delta_s\;\;\;\; \text{where } i = \{1,\dots,\mathcal{K}\}\:,\\
  \frac{\partial^2 \delta \bm{x}(s)}{\partial s^2} &\approx \frac{\delta \bm{x}(s_i+\Delta_s) + \delta \bm{x}(s_i-\Delta_s) - 2\delta \bm{x}(s_i)}{(\Delta_s)^2}\:,
\end{align}
\end{subequations}
and the matrices $D$ and $J$ turn into block matrices such that
\begin{equation}
  D =
  \begin{pmatrix}
  \left[D_1\right]_{\mathcal{K}\times \mathcal{K}} & 0 & 0 & \cdots\\
  0 & \left[D_2\right]_{\mathcal{K}\times \mathcal{K}} & 0 & \cdots\\
  \vdots & \vdots & \vdots & \ddots
  \end{pmatrix}.
\end{equation}
The $ij^{\text{th}}$ block of $J$ is $J_{ij}I_{\mathcal{K}\times \mathcal{K}}$ to give
\begin{equation}
  J = \begin{pmatrix}
  \left[J_{11}\right]_{\mathcal{K}\times \mathcal{K}} & \left[J_{12}\right]_{\mathcal{K}\times \mathcal{K}} & \cdots\\
  \left[J_{21}\right]_{\mathcal{K}\times \mathcal{K}} & \left[J_{22}\right]_{\mathcal{K}\times \mathcal{K}} & \cdots\\
  \left[J_{31}\right]_{\mathcal{K}\times \mathcal{K}} & \left[J_{32}\right]_{\mathcal{K}\times \mathcal{K}} & \cdots\\
  \vdots & \vdots & \ddots
\end{pmatrix}.
\end{equation}
Compartmentalisation results in a system of $n\mathcal{K}$ coupled SDEs
\begin{equation}
  d\delta \bm{x} = A\delta \bm{x}\:dt + \bm{\eta}\: dt\: ,
  \label{StochEq}
\end{equation}
with $A= D/\Delta_s^2 + J$ , $\langle\eta_i(t)\rangle = 0$ and $\langle\eta_i(t)\eta_j(t')\rangle = B_{ij}G_{ij}(t,t')$.

To calculate the power spectra of the system we introduce the discrete cosine transform \cite{Briggs1995}
\begin{equation}
  f_\kappa = \Delta_s\sum_{j=1}^\mathcal{K} \cos[\kappa\Delta_s(j-1)]f(s_j)\: .
\end{equation}
The cosine transform incorporates  the von Neumann (no flux) boundary conditions, which are commonly used for reaction-diffusion systems, however, other boundary conditions can easily be implemented \cite{Briggs1995}. Due to the boundary conditions we require $\kappa = m\pi/l$ with $m \in \{0,1,2,\dots\}$. We refer to $m$ as the {\em spatial mode}. Note that the use of $(j-1)$ instead of $j$ which is due to the fact that the compartment labelling starts at $j=1$. Hence, by applying the spatial Fourier transform we reduce the system of $n\mathcal{K}$ SDEs \eqref{StochEq} to a system of $n$ coupled SDEs
\begin{equation}
  d\delta \bm{x}_\kappa = A_\kappa\delta \bm{x}_\kappa\: dt + \bm{\eta}_\kappa\: dt\: .
  \label{StockEqSpat}
\end{equation}

Finally, applying the temporal Fourier transform
\begin{equation}
  \tilde{f}(\omega) = \int_{-\infty}^\infty f(t)e^{-i\omega t}dt\: ,
\end{equation}
we get an expression for $\delta\tilde{\bm{x}}_\kappa(\omega)$. Note that the Fourier transform always exists if we consider the system to have a fixed point \cite{Woolley2011b}.
\begin{equation}
  \delta\tilde{\bm{x}}_\kappa(\omega) = \left[-A_\kappa-i\omega\right]^{-1}\tilde{\bm{\eta}}_\kappa(\omega)\: .
\end{equation}
Therefore, the power spectra of the pattern, $P_{\delta x}(\kappa,\omega)$, can be expressed as the diagonal elements of
\begin{equation}
   \langle \delta \tilde{\bm{x}}_\kappa(\omega)\delta \tilde{\bm{x}}_\kappa^\dagger(\omega)\rangle = \Phi^{-1}\underbrace{\langle\tilde{\bm{\eta}}_\kappa(\omega)\tilde{\bm{\eta}}_\kappa^\dagger(\omega)\rangle}_{N} \left(\Phi^{-1}\right)^\dagger\: ,
\end{equation}
where we defined
\begin{equation}
  \Phi = -\left[A_\kappa + i\omega\right]\: ,
\end{equation}
and $\dagger$ denotes the hermitian conjugate of a matrix.
In order to be guaranteed real power spectra we need $N = N^\dagger$.
\bk In the case of Gaussian noise we have
\begin{equation}
  N = \mathcal{F}_\text{cos}\left(\mathcal{F}(BG(t,t'))\right),
\end{equation}
where $\mathcal{F}_\text{cos}$ denotes the cosine transform and $\mathcal{F}$ the temporal Fourier transform. Note that when all temporal correlations are equal such that $G_{ij}(t,t') = g(t,t')$ the transform factor becomes
\begin{equation}
  N = \mathcal{F}_\text{cos}\left(B\right)\mathcal{F}(g(t,t')) = B_\kappa P_\text{Correlations}(\omega)\: ,
\end{equation}
where $P_\text{Correlations}$ is the power spectrum of the correlation function $g(t,t')$ and $B_\kappa$ is the $n\times n$ matrix of covariances.
For white noise we take $g(t,t')=\delta(t-t')/dt$ to reduce \eqref{StochEq} to the It{\^o} form
\begin{equation}
  d\delta \bm{x} = A\delta \bm{x}\:dt + \bm{\eta}\: \sqrt{dt}\: ,
\end{equation}
with $\langle \eta_i(t)\eta_j(t')\rangle = B_{ij}\delta(t-t')$ and, therefore, $P_\text{Correlations} = 1$. Hence, we see that for white noise the power spectra are
\begin{equation}
  \langle \delta \tilde{\bm{x}}_\kappa(\omega)\delta \tilde{\bm{x}}_\kappa^\dagger(\omega)\rangle = \left[\Phi^{-1}\langle\tilde{\bm{\eta}}_\kappa(\omega)\tilde{\bm{\eta}}_\kappa^\dagger(\omega)\rangle \left(\Phi^{-1}\right)^\dagger\right] = \left[\Phi^{-1} B_\kappa \left(\Phi^{-1}\right)^\dagger\right] = P_\text{white}\: .
\end{equation}
For general temporal correlations $g(t,t')$ we have
\begin{equation}
  \langle \delta \tilde{\bm{x}}_\kappa(\omega)\delta \tilde{\bm{x}}_\kappa^\dagger(\omega)\rangle = \left[\Phi^{-1} B_\kappa P_\text{Correlations}\left(\Phi^{-1}\right)^\dagger\right] = P_\text{white}P_\text{Correlations}\: ,
\end{equation}
such that the spectra of the noise colour appear as a multiplicative factor modulating the white noise spectra.

\subsection{Application to the Schnakenberg System}

We illustrate our analysis via the example of the Schnakenberg kinetics. The Schnakenberg system is an $n=2$ species reaction-diffusion system which, despite its apparent simplicity, exhibits a wealth of different behaviours in the deterministic \cite{Iron2004,Maini2012,Ward2002,Flach2007} as well as stochastic \cite{Schumacher2013,Woolley2011a} regimes. The Schnakenberg system is a fully open system and, hence, both species can be subject to stochastic inflow.
The system follows the reaction scheme
\begin{equation}
  X_2 \xleftarrow{k_2}\emptyset \xrightleftharpoons[k_{-1}]{k_1} X_1\: ,\qquad\qquad\qquad   2X_1+X_2 \xrightarrow{k_3} 3X_1\:.
  \label{Schnakenberg}
\end{equation}

The dynamics of the Schnakenberg system is described by the system of PDEs
\begin{align}
  \frac{\partial x_1}{\partial t} &= D_1\frac{\partial^2 x_1}{\partial s^2} +k_1 - k_{-1} x_1 + k_3x_1^2x_2 ,\nonumber\\
  \frac{\partial x_2}{\partial t} &= D_2\frac{\partial^2 x_2}{\partial s^2} + k_2 - k_3x_1^2x_2 .
\end{align}

First we check for a potential multistationarity in the Schnakenberg system by formulating the matrices $m_1$ and $m_2$.
\begin{equation}
  m_1 = \begin{pmatrix}
  1 & -1 & 0 & 1\\
  0 & 0 & 1 & -1
\end{pmatrix}
  \quad\quad\text{and}\quad\quad m_2 = \begin{pmatrix}
  0 & 0\\
  1 & 0\\
  0 & 0\\
  2 & 1
\end{pmatrix}.
\end{equation}
It is easy to check that $\text{ker}(m_2) = \{0\}$. Further, we have
\begin{equation}
  \sigma(\text{ker}(m_1)) = \{(+,+,0,0)^T,(-,0,+,+)^T\}\;\;\text{and}\;\; \sigma(\text{im}(m_2)) = \{(0,+,0,+)^T,(0,0,0,+)^T\}\:.
\end{equation}
Hence, applying the injectivity criterion \eqref{Inject} we see that the Schnakenberg system is monostationary for all positive parameter values and, therefore, stochastic fluctuations in the inflow parameters cannot trigger stochastic switching.

Next, we consider stochastic inflows which result in $k_i \to k_i + 1/\sqrt{\Omega}\eta_i$ for $i \in \{1,2\}$ and the equations
\begin{align}
  \frac{\partial x_1}{\partial t} &= D_1\frac{\partial^2 x_1}{\partial s^2} +k_1 - k_{-1} x_1 + k_3x_1^2x_2 + \frac{1}{\sqrt{\Omega}}\eta_1,\nonumber\\
  \frac{\partial x_2}{\partial t} &= D_2\frac{\partial^2 x_2}{\partial s^2} + k_2 - k_3x_1^2x_2 + \frac{1}{\sqrt{\Omega}}\eta_2.
  \label{SchnakStoch}
\end{align}

Linearising equations \eqref{SchnakStoch} about the steady state $x^*_1 = (k_1 + k_2)k^{-1}_{-1}$, $x^*_2 = k_2k_{-1}^2k_3^{-1}(k_1+k_2)^{-2}$ and discretising space we get a system of linear stochastic differential equations similar to the ones arising from the study of internal noise \cite{Schumacher2013}
\begin{align}
d\bm{x} &= \bm{Ax} + \bm{\eta},
\label{SchnakLin}\\
\intertext{with}
A &= \begin{pmatrix}
  \bm{a} & \bm{b}\\
  \bm{c} & \bm{d}
\end{pmatrix},
\end{align}
where $\bm{a},\bm{d}$ are tridiagonal matrices with diagonal elements
\begin{align}
  a_0 &= -2D_1/\Delta_s^2 -k_{-1}+2k_3x_1^*x_2^*,\nonumber\\
  d_0 &= - 2D_2/\Delta_s^2 - k_3x_1^{*2},\\
  \intertext{and off-diagonal (sub- and super-diagonal) elements}
  a_1 &= D_1/\Delta_s^2 \equiv d_u,\nonumber\\
  d_1 &= D_2/\Delta_s^2 \equiv d_v.
\end{align}
The matrices $\bm{b}$ and $\bm{c}$ are diagonal matrices with entries
\begin{align}
  b_0 &= k_2x_1^{*2},\nonumber\\
  c_0 &= -2k_3x_1^*x_2^*.
\end{align}

Note that the vector $\bm{x}$ has $2\mathcal{K}$ components. There are $\mathcal{K}$ components for species $x_1$ and $\mathcal{K}$ components for species $x_2$.

Hence, Fourier transforming equation \eqref{SchnakLin} in space and time we can compute the power spectra,
\begin{align}
  P(k,\omega) &= \left[A_\kappa+i\omega\right]^{-1}\langle \tilde{\eta}_\kappa(\omega)\tilde{\eta}_\kappa^\dagger(\omega)\rangle\left(\left[A_\kappa+i\omega\right]^{-1}\right)^\dagger,\\
  \intertext{with}
  A_\kappa &= \begin{pmatrix}
  a_0 + 2a_1\cos{k\Delta_s} & b_0\\
  c_0 & d_0+2d_1\cos{k\Delta_s}
  \end{pmatrix}.
\end{align}

In the following section we will discuss the effect of various noise correlations on the power spectra, and hence, the patterns generated.

\section{Computational Results}\label{Computations}

In this section we highlight the computational patterns generated by a Turing system with coloured noise.
In subsection \ref{NumericalMethods} we start with a brief discussion of the numerical methods used before we discuss the differences between parameter noise as considered in this paper and internal noise in subsection \ref{WhiteNoiseSection}. We then review systems with white noise which were mainly derived in \cite{Schumacher2013} and in the following subsections we proceed to relax the assumption of uncorrelated noise to see the effect of noise colouring. This section is divided into two parts, the first one treating stochastic processes which may arise from experimental set-ups, such as exponential broadening of the noise correlation (Ornstein-Uhlenbeck noise) or excitation of the zero frequency modes. The second part will treat sources of noise due to stochastic subnetworks and noise mixtures. An overview of our main results is given in Table \ref{TabResults}

\begin{flushleft}
\begin{table}[h!!!]
\resizebox{0.9\textwidth}{!}{\begin{minipage}{\textwidth}
\begin{tabular}{|l|l|l|l|l|}
\hline
\textbf{Colour} & \textbf{Correlation Function} & \textbf{Effect} & \textbf{Possible Origin}& \textbf{Section}\\
\hline
White & $1$ & N/A & internal& \ref{WhiteNoiseSection}\\
Ornstein-Uhlenbeck & $1/\left(\omega^2\tau^2+1\right)$ & suppresses small $\omega$ & experimental apparatus& \ref{OrUlSection}\\
Red & $1/\omega^2$ & excites small $\omega$ & deterministic & \ref{RedSection}\\
Predator-Prey & $\left(\alpha+\beta\omega^2\right)/\left(\omega^4 + \gamma\omega^2+\delta \right)$ & induces oscillations & stochastic subsystems & \ref{OscSection}\\
\hline
\end{tabular}
\end{minipage}}
\caption{An overview of our main results on various noise colours.}
\label{TabResults}
\end{table}
\end{flushleft}

\subsection{Numerical Methods}\label{NumericalMethods}

The system of SDEs \eqref{StochEq} is simulated by an Euler-Maruyama scheme \cite{Kloeden1995} with time step $\Delta t$ such that
\begin{equation}
    \bm{x}(t+\Delta t) = \bm{x}(t) + \bm{Ax}\Delta t + \bm{\eta}\Delta t.
\end{equation}
The important step of the integration is to find the correct vector $\bm{\eta}$.

The white noise and Ornstein-Uhlenbeck noise are generated by an auxiliary noise process. In particular, at teach time step the white noise is sampled from a multivariate Gaussian distribution $$\bm{\eta}\Delta t \sim \mathcal{N}(0, \bm{B}\Delta t).$$ The Ornstein-Uhlenbeck process is a generated by the stochastic differential equation $$\frac{d\bm{\eta}}{dt} = -\frac{1}{\tau}\bm{\eta} + \frac{\sqrt{\bm{B}}}{\tau}\bm{\xi},$$ where $\bm{\xi}$ is a standard white noise vector. Therefore, at time $t$ the vector $\bm{\eta}(t)$ is added to the to the system of SDEs. The Ornstein-Uhlenbeck ``auxiliary equation'' is integrated by an Euler-Maruyama scheme as described in \cite{Milshtein1994}.

To simulate power law noise for which, in general, no auxialiary SDE exists we use inverse transforms \cite{Timmer1995}. To generate a vector $\bm{\eta}(t)$ with correlations $\langle\bm{\eta}(t)\bm{\eta}^T(t')\rangle = \bm{B}g(t-t')$, we first use the fact that $\bm{\eta} = \sqrt{\bm{B}}\bm{\xi}$ where $\bm{\xi}$ may be correlated in time but not in space, i.e. $\langle\bm{\xi}(t)\bm{\xi}^T(t')\rangle = \delta_{ij} g(t-t')$. Then, let the power spectrum of $\xi_i$ be $1/\omega^\alpha$ as described in subsection \ref{RedSection} and use the algorithm of \cite{Timmer1995} to create a time series. Multiplication of $\bm{\xi}(t)$ with $\sqrt{\bm{B}}$ gives the desired noise process, $\bm{\eta}(t)$, which can be added to the SDE system.

Auxiliary networks as in \ref{OscSection} can be simulated by either method and in this paper the auxiliary network input is generated by using the inverse transforms technique.

\subsection{White Noise}\label{WhiteNoiseSection}

First, we consider external white noise. In this special case the noise vector $\bm{\eta}$ is just a Wiener process with correlation matrix $B$. This case is mathematically identical to the case of internal noise in the weak noise limit as studied in \cite{Schumacher2013}. The main differences between internal noise and the parameter noise considered in this paper are the amplitude of the noise and the exact forms of the covariances of the stochastic processes $\bm{\eta}$.

Both approximations (the weak noise limit, as well as our ``truncated Gaussian'' approximation) assume that the stochastic effect is a perturbation to the deterministic limit. However, as derived in \cite{Schumacher2013}, the covariance matrix of the stochastic processes $\bm{\eta}$ is determined by the microscopic processes whereas for external noise, whose origin can be manifold, the covariance matrix is be arbitrary. For mathematical simplicity we choose the covariance matrix
\begin{equation}
  B_\kappa = \begin{pmatrix}
  1 & 1\\
  1 & 1
  \end{pmatrix}.
\end{equation}

Throughout the remainder of this paper we fix the parameter values to
\begin{align}
  k_1 &= k_2 = 10.0,\nonumber\\
  k_3 &= 0.01,\nonumber\\
  k_{-1} &= 20.0,\nonumber\\
  \intertext{with diffusion coefficients}
  D_1 &= 10^{-4},\nonumber\\
  D_2 &= 10^{-2},
  \label{ParamVals}
\end{align}
 which results in steady state concentrations of $x_1^* = 1$, $x_2^* = 1000$. The parameters chosen are not generic, but they represent a particularly interesting point in parameter space. The parameters are in the (stable) oscillatory regime of the Schnakenberg system and outside the Turing space. The main function of the noise will be to temporarily move the system into the Turing regime. Note that the system has large inflows (and outflows) compared to the chemical reaction parametrised by $k_3$ which allows for larger variations in the noise translating into larger pattern amplitudes. The modifications to the power spectra due to coloured noise are generic and apply all points in parameter space. However, the visibility of these modifications depends on the point in parameter space and the correlation matrix.  Similarly, the results on amplitude of the patterns are parameter-dependent. To simulate the system, we discretise the space into $40$ compartments of width $\Delta_s = 0.0025$ (which gives a total domain length of $L = 0.1$).

\begin{figure}
\centering
 \subfloat[The deterministic solution of the Turing system at the given parameter point\label{fig1:det}]{\includegraphics[width=.5\textwidth]{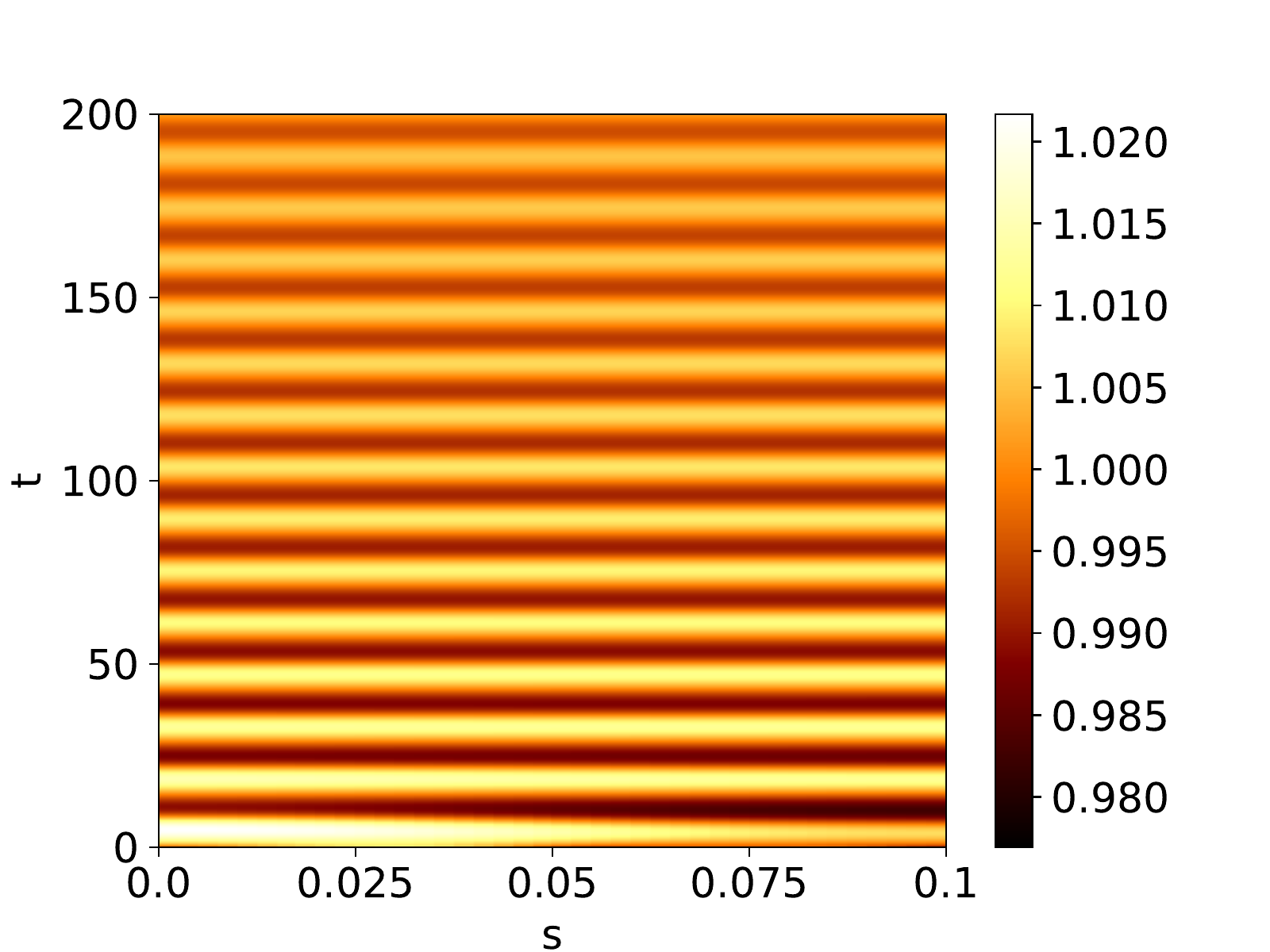}}
 \subfloat[A stochastic realisation with white noise\label{fig1:stoch}]{\includegraphics[width=.5\textwidth]{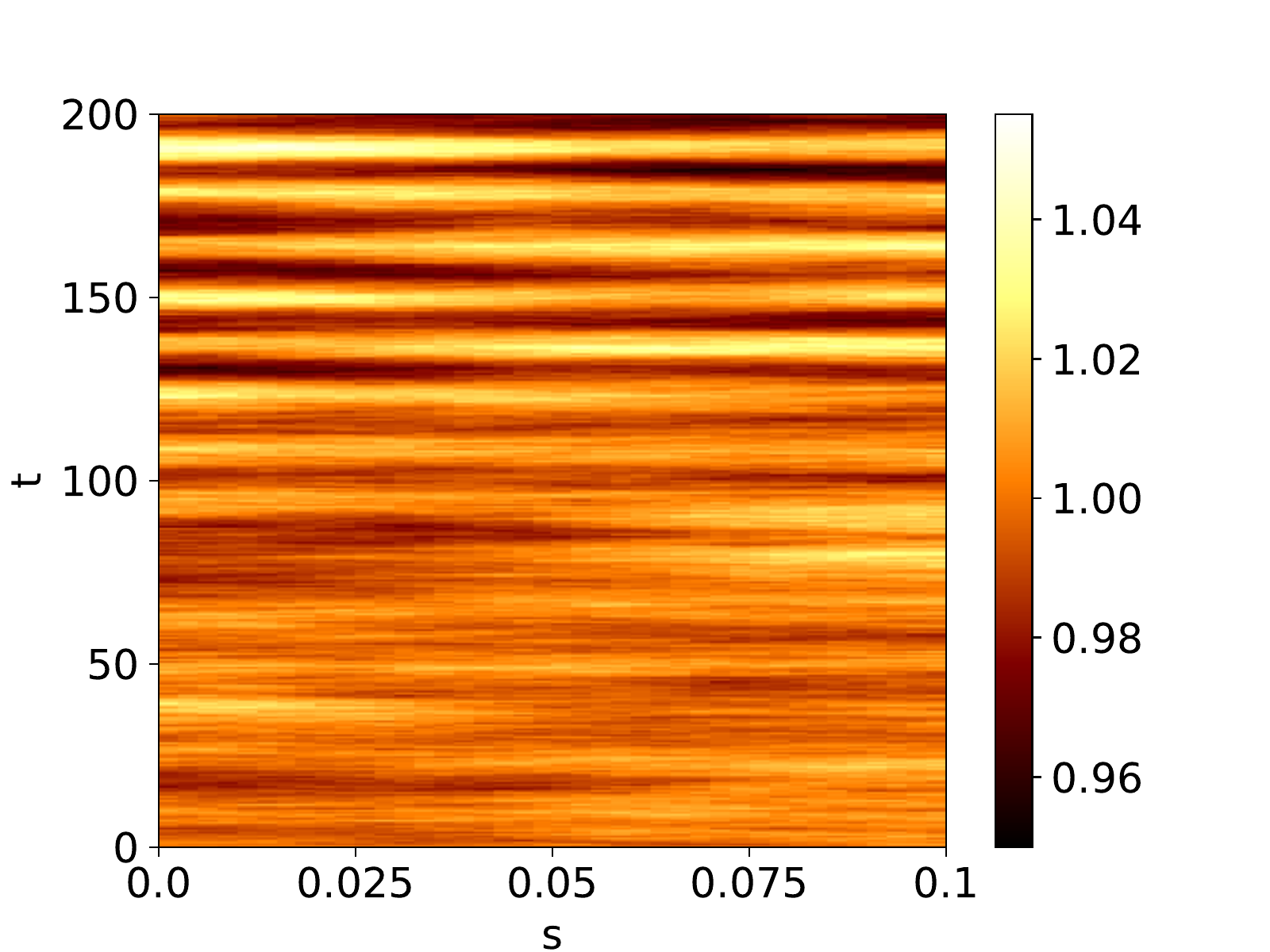}}\newline
 \subfloat[The power spectra of the analytical calculations (dashed lines) and their simulated curves averaged over 50 realisations of length $T=1000$ and subsystem size $\Omega=5000$. We attribute the differences in peak height to the finite time steps used.\label{fig1:power}]{\includegraphics[width=.7\textwidth]{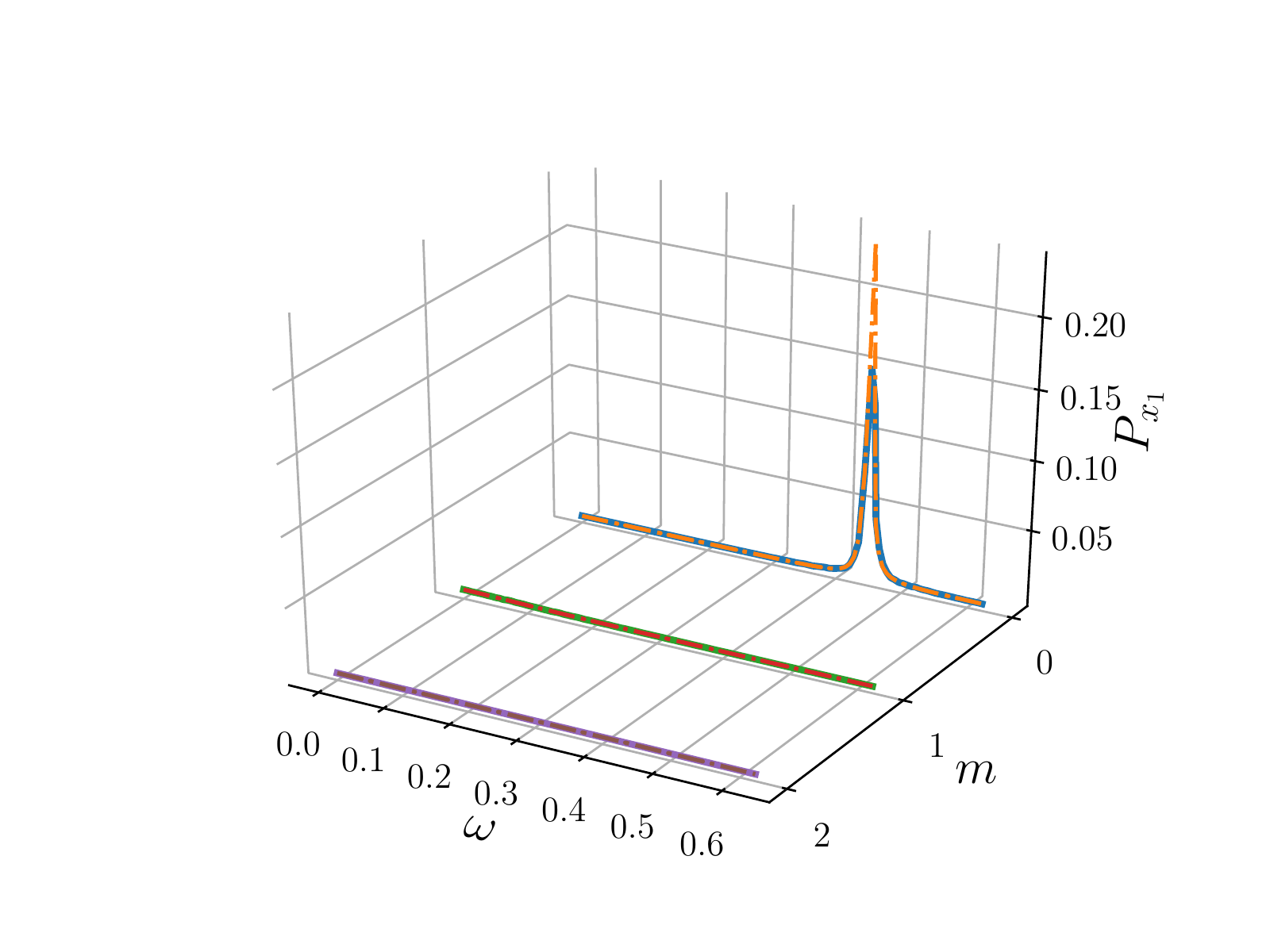}}
  \caption{An overview of the deterministic and white noise behaviour of the system with parameters as in \eqref{ParamVals}. The peak in \ref{fig1:power} at zero spatial mode, $m=0$, indicates temporal oscillations which result from deterministic limit cycle oscillations.  Due to the choice of reaction kinetics the temporal oscillations of species $x_1$ and $x_2$ are in phase.}
  \label{White}
\end{figure}

Simulating equation \eqref{SchnakLin} under the influence of white noise and with the given parameter values we obtain Figure \ref{White} and its corresponding averaged power spectrum.
The power spectrum shows the temporal frequencies, $\omega$, and spatial modes, $m$, present in the pattern. The amplitude of the oscillations is about $3\%$ of the steady state value for species $x_1$ whereas for species $x_2$ it is much lower. Therefore, we assume that any patterning of $x_2$ is not generally measurable and, therefore, we focus our attention on $x_1$. The prominent temporal oscillations with no spatial dependency manifest themselves as a sharp peak in the power spectrum. This is due to the fact that the chosen parameter point is in the oscillatory regime of the Schnakenberg model. Upon close inspection slight spatial variations in the pattern of $x_1$ can be seen, but with white noise these are not very pronounced. In the following sections we show how noise colour can enhance spatial modes, create additional oscillations or even create a stable pattern.

\subsection{Ornstein-Uhlenbeck Noise}\label{OrUlSection}

Next, we investigate the effect of exponentially correlated noise also known as Ornstein-Uhlenbeck noise \cite{Hanggi1995}. This stochastic process has a finite correlation time $\tau$ which we interpret as a response time. Consider an experiment in which the inflow rate is highly sensitive to the ambient temperature. Further, suppose this temperature undergoes random fluctuations about a regulated mean value. Therefore, the correlation time $\tau$ could represent the average response time of the temperature regulator. The correlation time $\tau$ could also represent an intrinsic relaxation time of the system such as found in motility induced phase separation of bacteria \cite{Cates2015}.

\begin{figure}
 \centering
 \subfloat{\includegraphics[width=.5\textwidth]{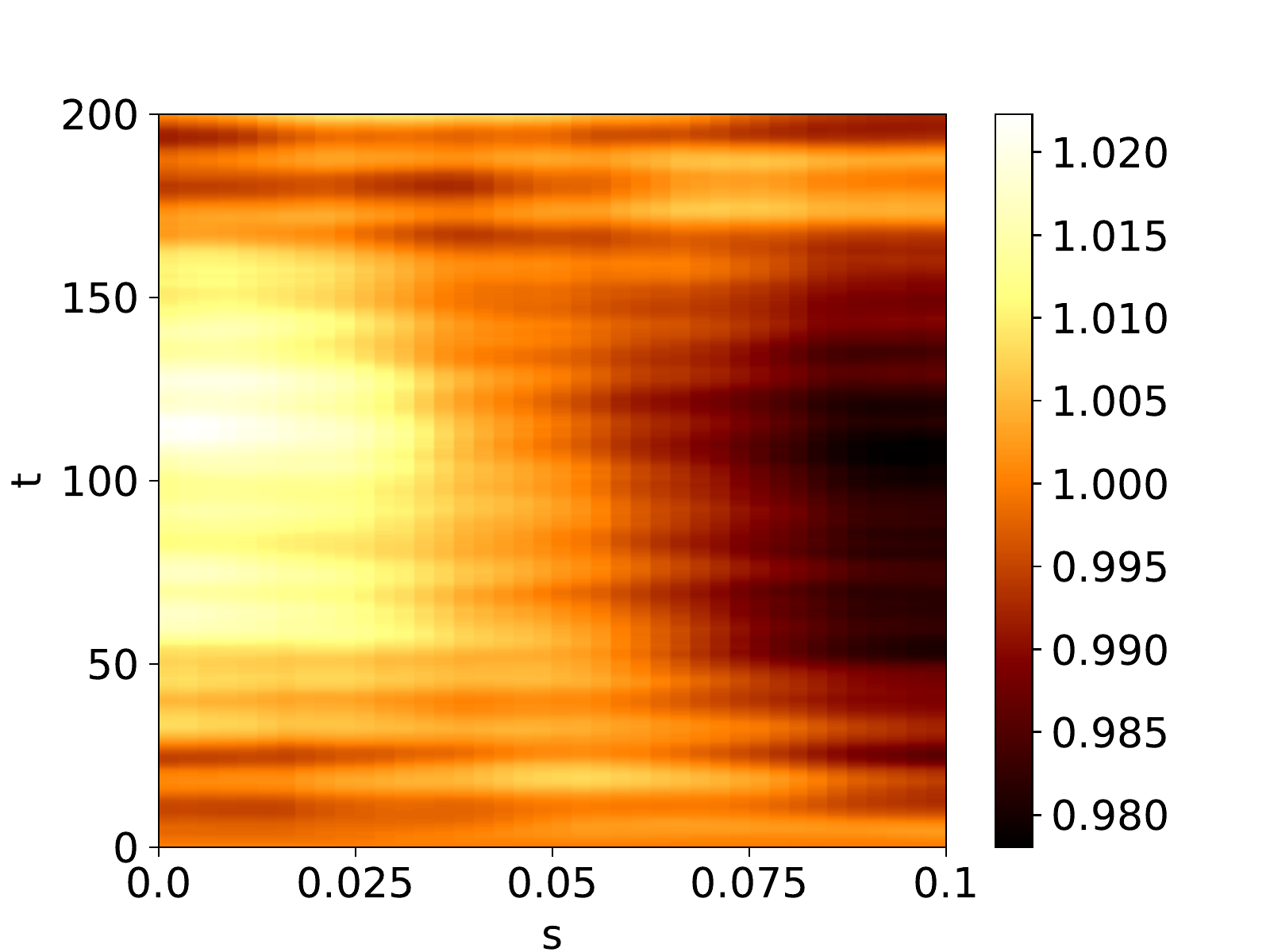}} 
 \subfloat{\includegraphics[width=.5\textwidth]{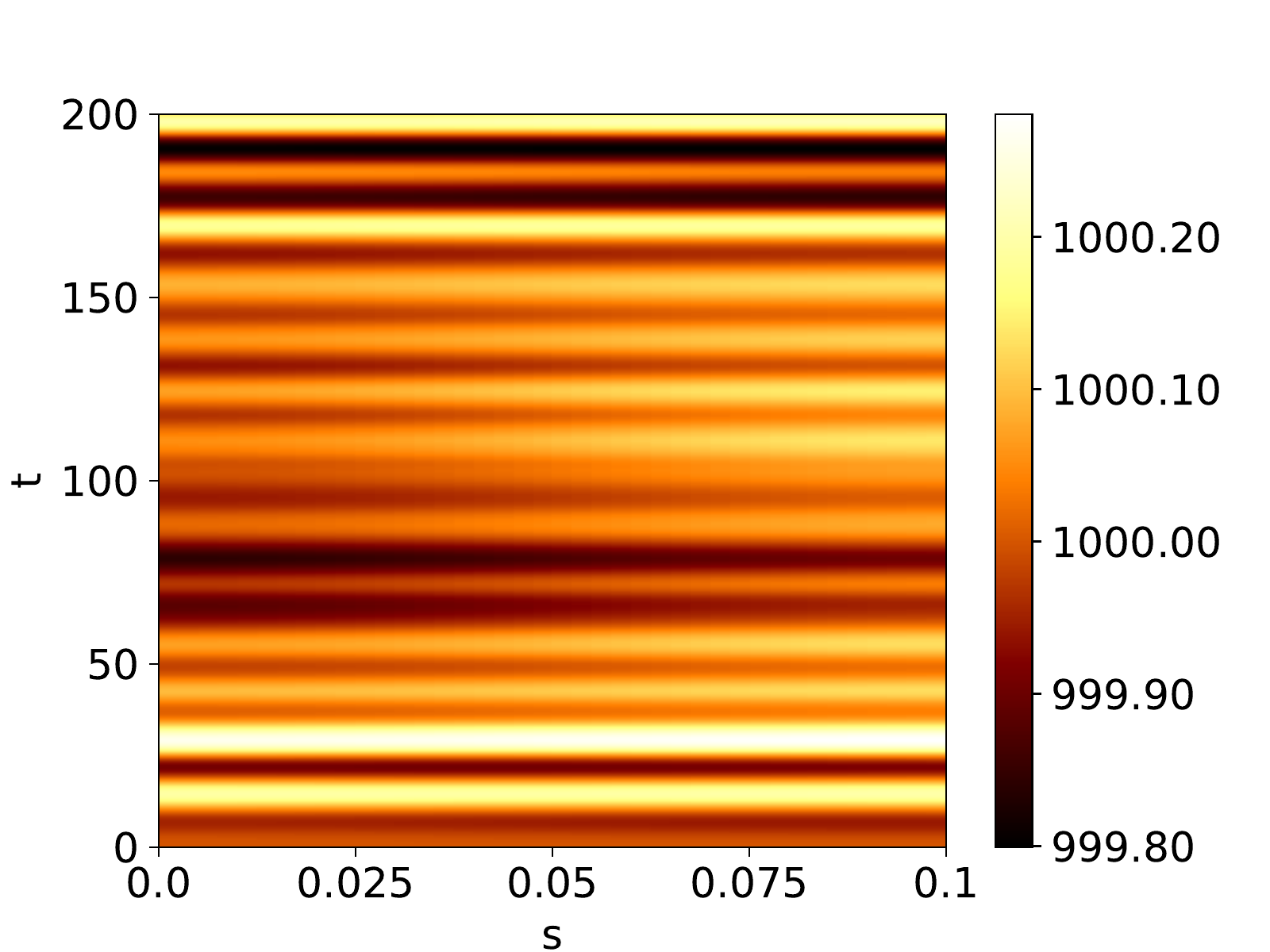}}\newline 
 \subfloat{\includegraphics[width=.47\textwidth]{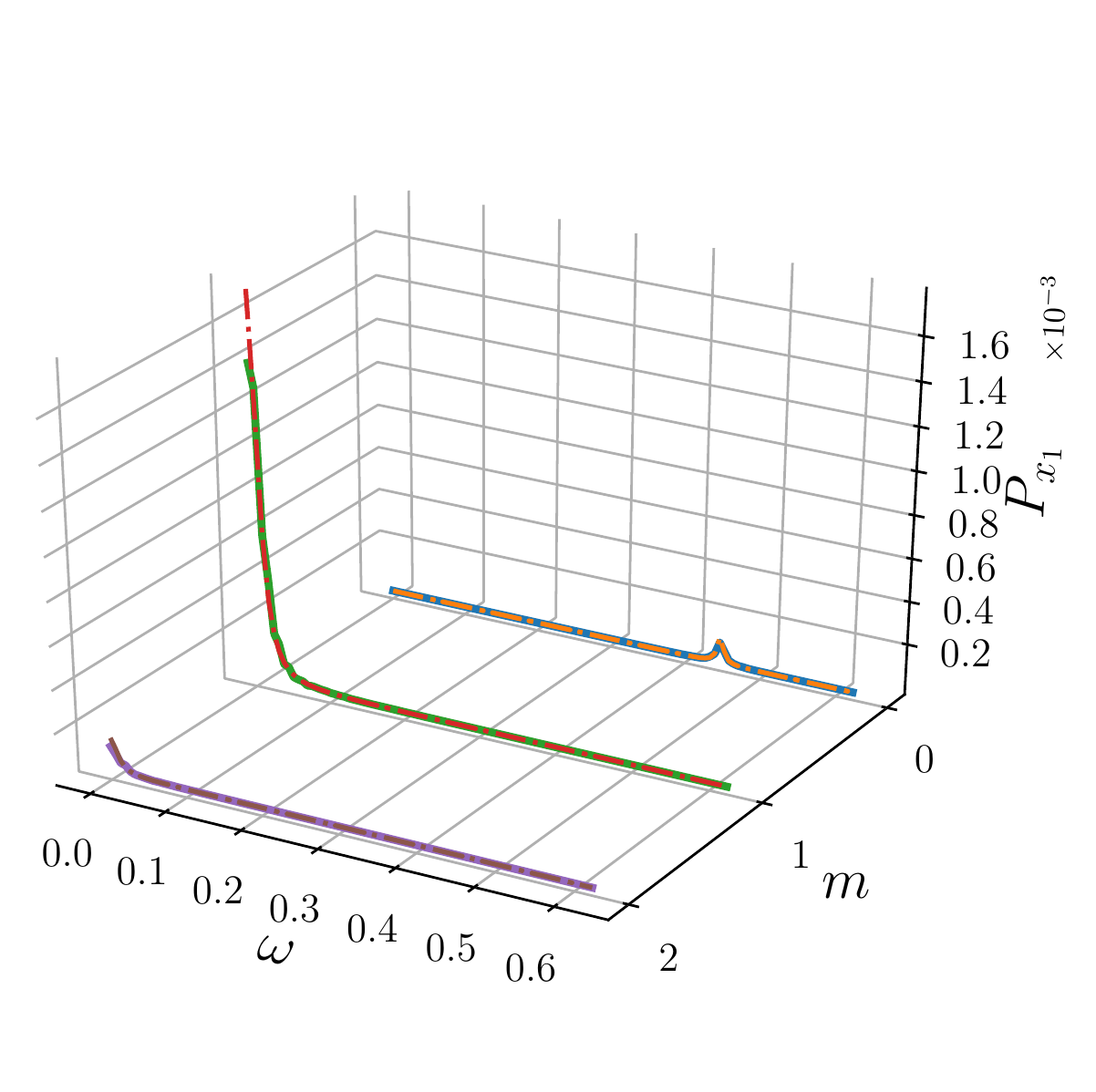}} 
 \subfloat{\includegraphics[width=.47\textwidth]{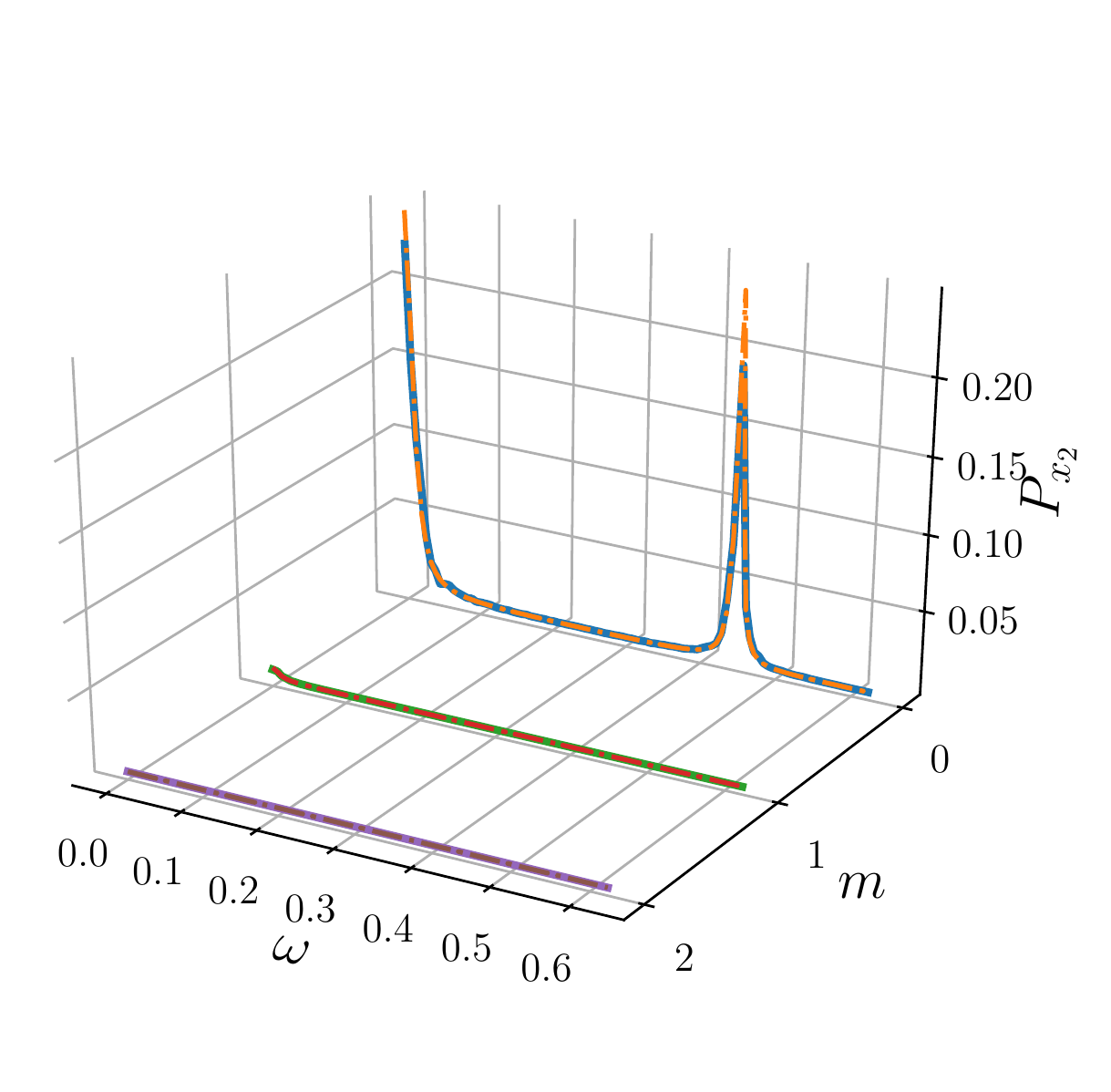}} 
 \caption{The patterns and power spectra for the species $x_1$ (left) and $x_2$ (right) with noise generated by an Ornstein-Uhlenbeck process and $\tau = 100$. Spatial patterns become visible in the Ornstein-Uhlenbeck case which can be attributed to its dampening effect on temporal oscillations. This can be observed as a visible excitation of the $m=1$ mode in the power spectrum of $x_1$ with the dashed line representing the analytical prediction. The well-known phenomenon of polarity switching \cite{Schumacher2013} is observed in the pattern of $x_1$. The spectra were averaged over 50 repetitions with $T=1000$ and subsystem size $\Omega = 100$.}
 \label{OrUl}
\end{figure}

We make the simplification that all the temporal correlations in equation \eqref{Correlations} originate from Ornstein-Uhlenbeck processes with the same correlation time, $\tau$, such that
\begin{equation}
  g_{ij}(t,t') = g(t,t') = \frac{1}{2\tau}e^{-\tfrac{|t-t'|}{\tau}}.
\end{equation}
Hence, it follows that
\begin{equation}
  \langle\eta_\kappa(\omega)\eta_\kappa(\omega)^\dagger\rangle = B_\kappa \frac{1}{\omega^2\tau^2 + 1}
\end{equation}
where $B_\kappa$ is the same matrix as in the white noise case and therefore $P(k,\omega)_\text{Ornstein-Uhlenbeck} = P(k,\omega)_\text{white}/(\omega^2\tau^2 +1)$.
Hence, the noise colouring will dampen temporal frequencies of $\omega \neq 0$ and, therefore, stabilise the pattern. Consequentially, in systems with Ornstein-Uhlenbeck noise we would expect any present stationary patterns that may be obscured by transient noise effects to be more prominent. We simulated the Schnakenberg system with Ornstein-Uhlenbeck noise and plotted the resulting patterns and power spectra in Figure \ref{OrUl}. The pattern of $x_2$ has the same characteristic as in the white noise case, except for a smaller amplitude. Interestingly, $x_1$ shows a very different behaviour to the white noise case. Clear spatial structures can be seen, in particular the phenomenon of polarity switching \cite{Woolley2011a,Schumacher2013}. A Turing pattern of mode $m=1$ is generated with a given polarity, namely, a minimum at $s=0$ and a maximum at $s=L$ or vice versa. Polarity switching describes the inherently stochastic phenomenon of a sudden change in polarity as can be observed in Figure \ref{OrUl}. The temporal oscillations, although still present, are reduced to a practically unobservable level. We attribute this to the fact that the exponential noise correlations dampen the oscillations present in the white noise system.

\subsection{Power Law Noise}\label{RedSection}

By power law noise we mean a stochastic process whose frequency distribution follows a power law in Fourier space
\begin{equation}
  P_\text{power} = \frac{1}{\omega^\alpha}.
\end{equation}
Various types of power law noise are common in engineering and are defined by colours summarised in Table \ref{ColoursTab}. In this paper we study red noise, white noise (subsection \ref{WhiteNoiseSection}) and violet noise to illustrate the effects of a positive, zero and negative exponent, $\alpha$.
\begin{table}[h!!!]
\begin{center}
\begin{tabular}{|l|r|}
  \hline
  \textbf{Colour} & $\alpha$\\
  \hline
  Red & $\alpha = 2$\\
  Pink & $\alpha=1$\\
  White & $\alpha = 0$\\
  Blue & $\alpha=-1$\\
  Violet & $\alpha = -2$\\
  \hline
\end{tabular}
\end{center}
\caption{The various colours of power law noise. In this paper we focus on red and violet noise.}
\label{ColoursTab}
\end{table}

\begin{figure}
    \centering
    \subfloat{\includegraphics[width=.48\textwidth]{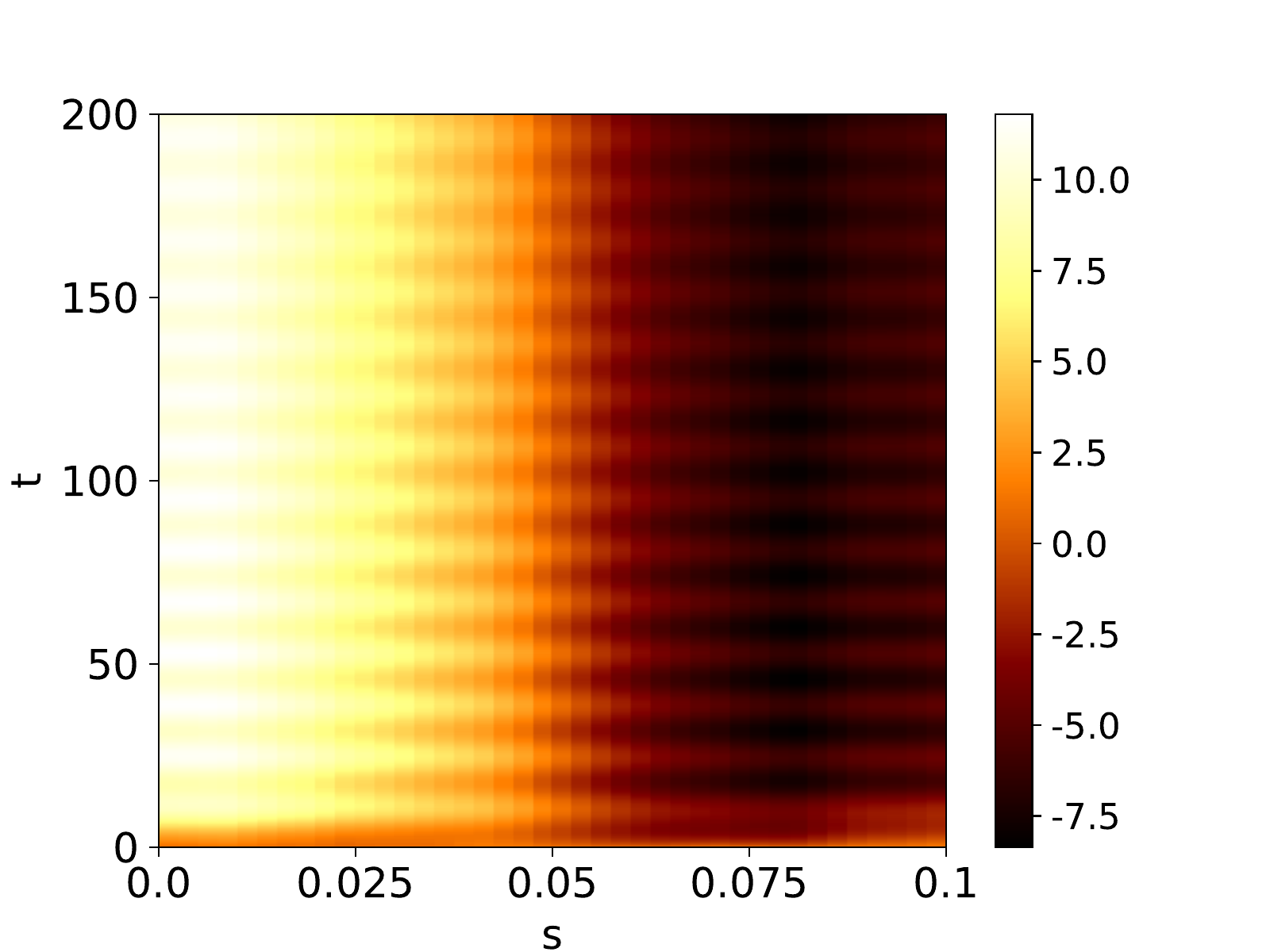}}
    \subfloat{\includegraphics[width=.48\textwidth]{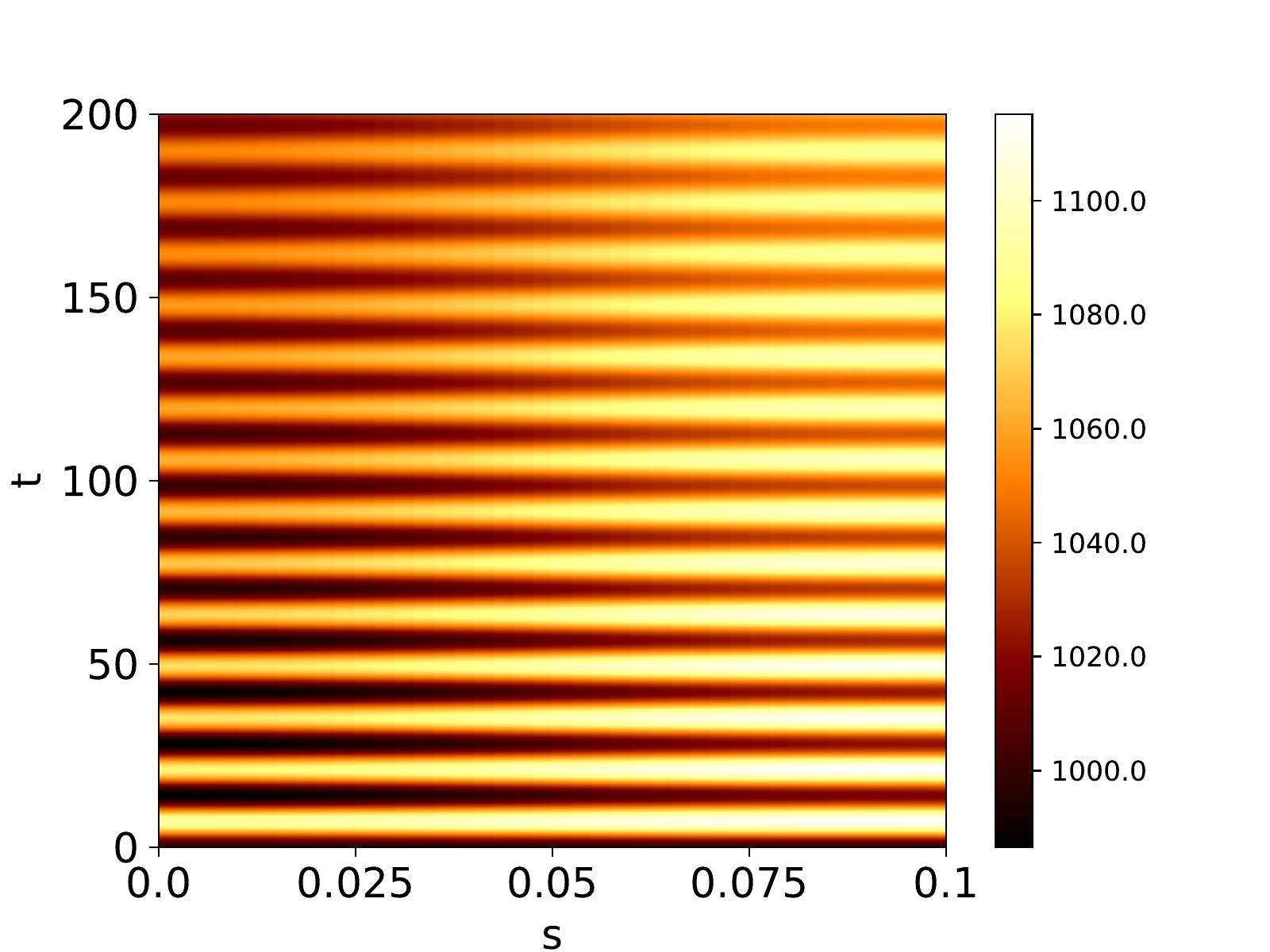}}\newline
    \subfloat{\includegraphics[width=.48\textwidth]{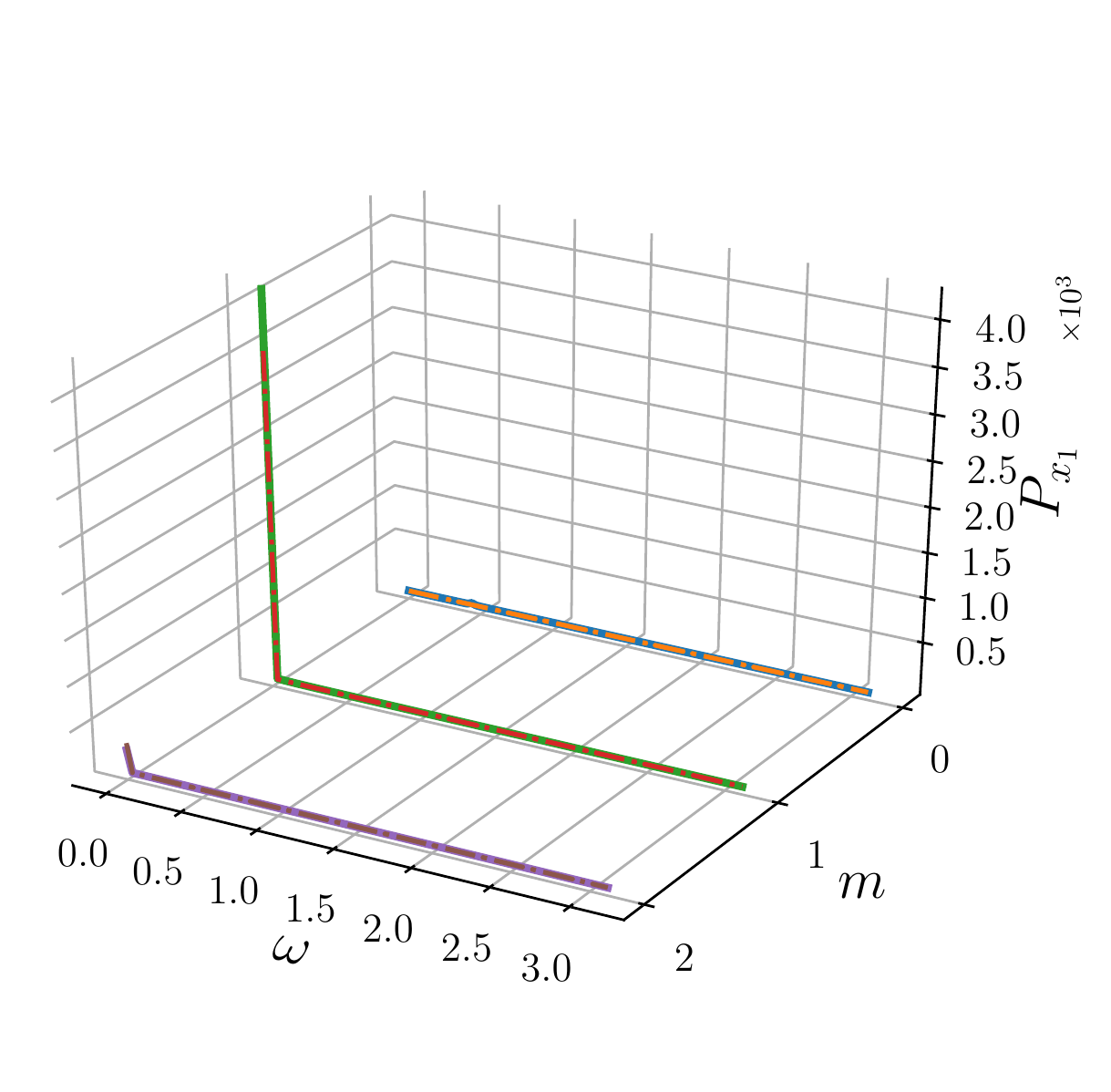}}
    \subfloat{\includegraphics[width=.48\textwidth]{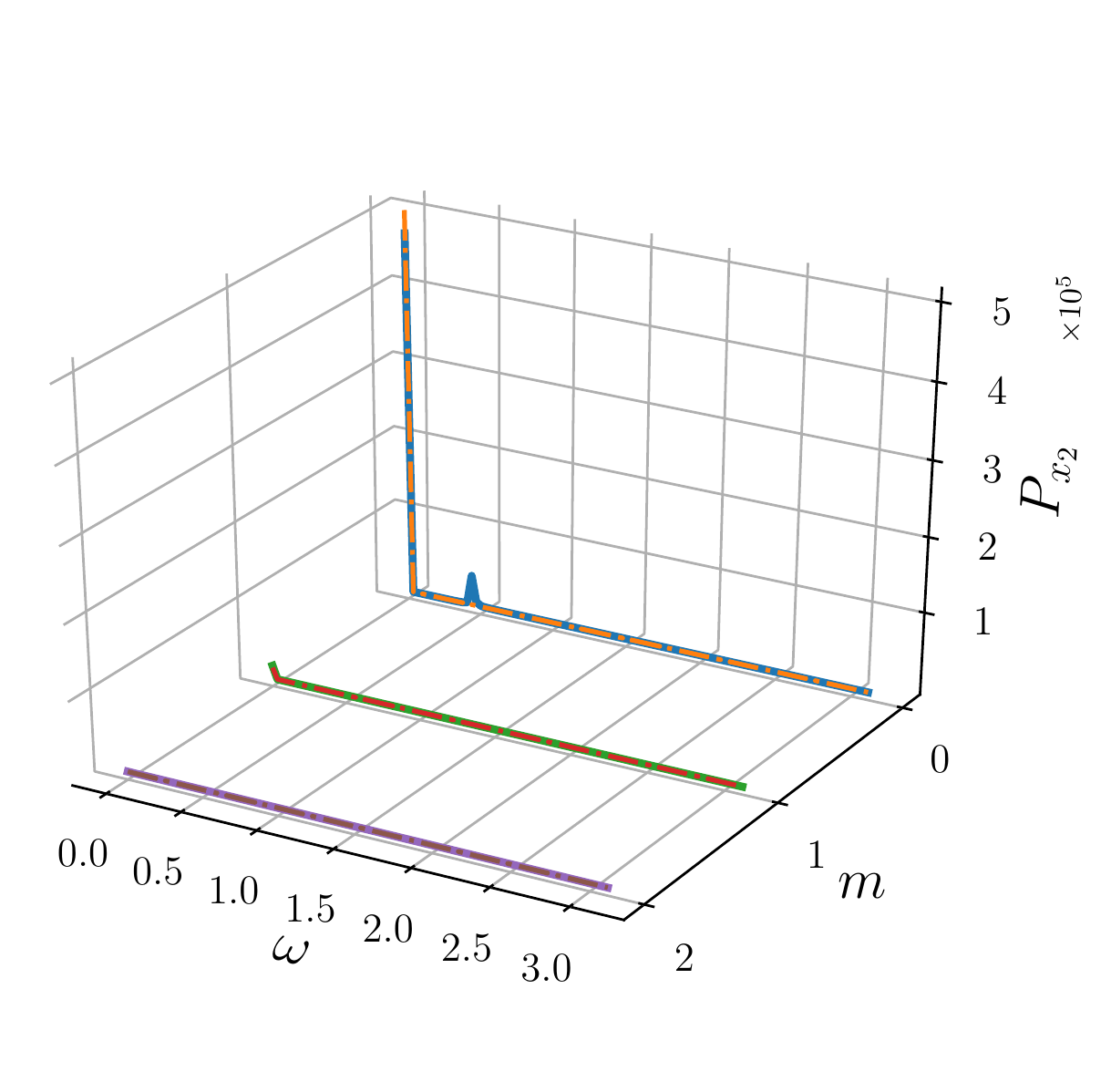}}
  \caption{The patterns and power spectra for the species $x_1$ (left) and $x_2$ (right) with noise generated by a red noise process. As predicted, the spectra are dominated by the behaviour the $\omega = 0$, which amounts to the stabilisation of a particular spatial mode. The species concentrations, however, go negative indicating that a full non-linear model needs to be used. The peak in the power spectrum for $x_2$ is a numerical artefact. The spectra were averaged over 50 repetitions with simulation time $T=200$. The subsystem size was $\Omega = 5000$.
  }
  \label{Pink}
\end{figure}

Red noise amplifies small temporal frequencies with $1/\omega^2 \to \infty$ as $\omega\to 0$. Therefore, the zero frequency behaviour dominates the pattern. As can be seen in Figure \ref{Pink}, due to the amplification of small $\omega$, the pattern becomes stable, even though the amplitude grows beyond the biologically viable (non-negative) bound indicating that non-linear effects need to be considered in the model.
To investigate this further, we look at the noise vector $\bm{\eta}$. For the simulation times chosen, $T=200$, the noise vector is independent of the time $t$ such that each $\eta_i$ has a constant, but random value. Therefore, the effect of red noise is a deterministic one and to get an accurate description of the system behaviour a full non-linear, deterministic compartment model with generic (random) inflow parameters needs to be analysed. The red noise case is similar to models which assume a cell-to-cell variability where a parameter is perturbed by a constant, sampled from a given distribution \cite{Lenive2016,Filippi2015}.
In contrast to conventional Turing theory in which the inflow parameter is constant throughout the domain, our analysis indicates that randomly varying inflows across the domain might actually facilitate pattern formation.

When investigating the power spectra we see that for species $x_1$ the simulations match the mathematical prediction. For $x_2$, however, we see that the peaks at the deterministic oscillation frequency have different heights. This indicates that either the oscillation amplitude is underestimated by the analytical prediction or, vice versa, the simulations exaggerate the oscillation amplitudes due to numerical limitations. Due to the agreement of prediction and simulations at all other points we disregard the different peak heights as numerical artefacts.

\begin{figure}
  \centering
    \subfloat{\includegraphics[width=.48\textwidth]{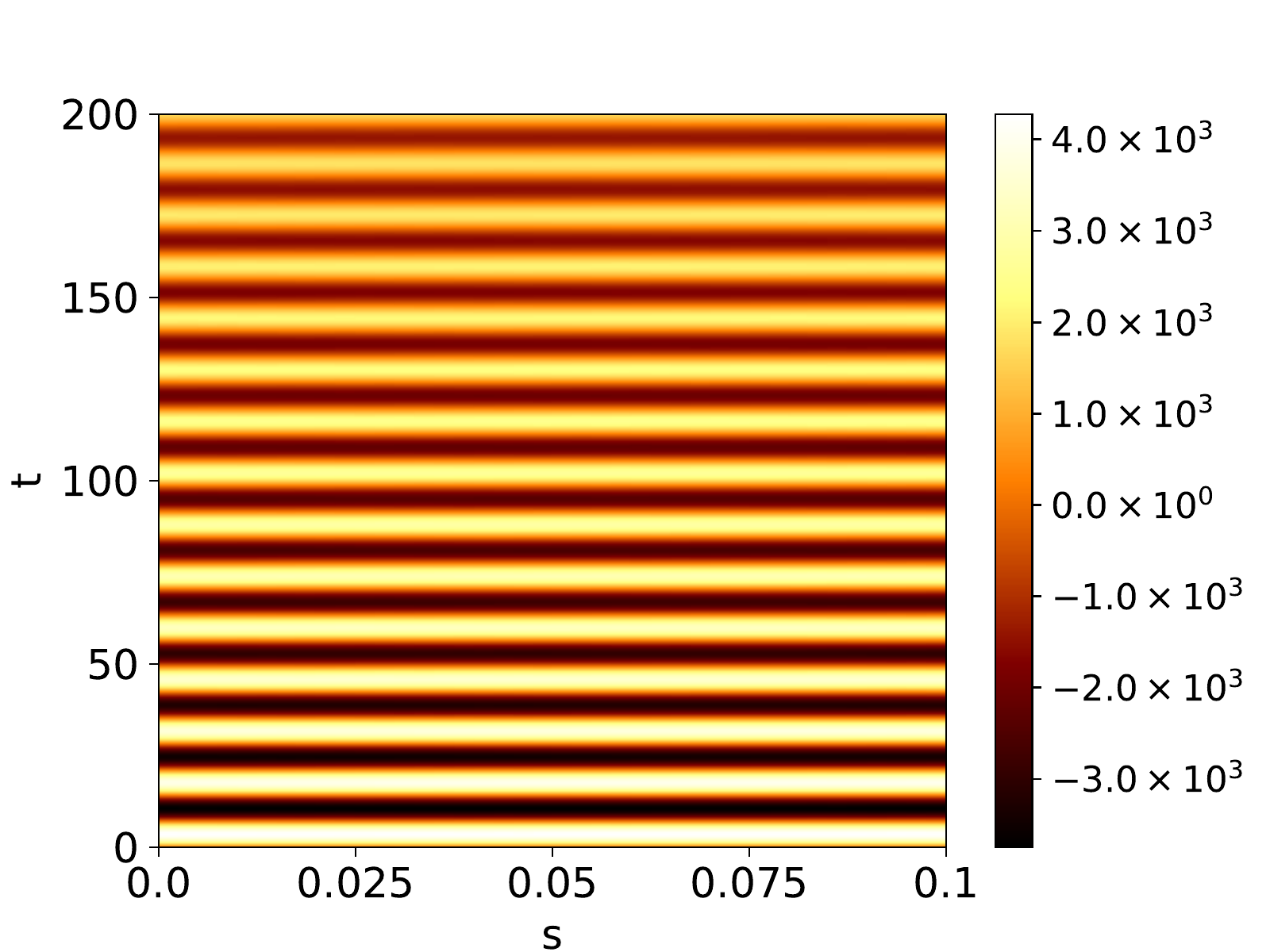}}
    \subfloat{\includegraphics[width=.48\textwidth]{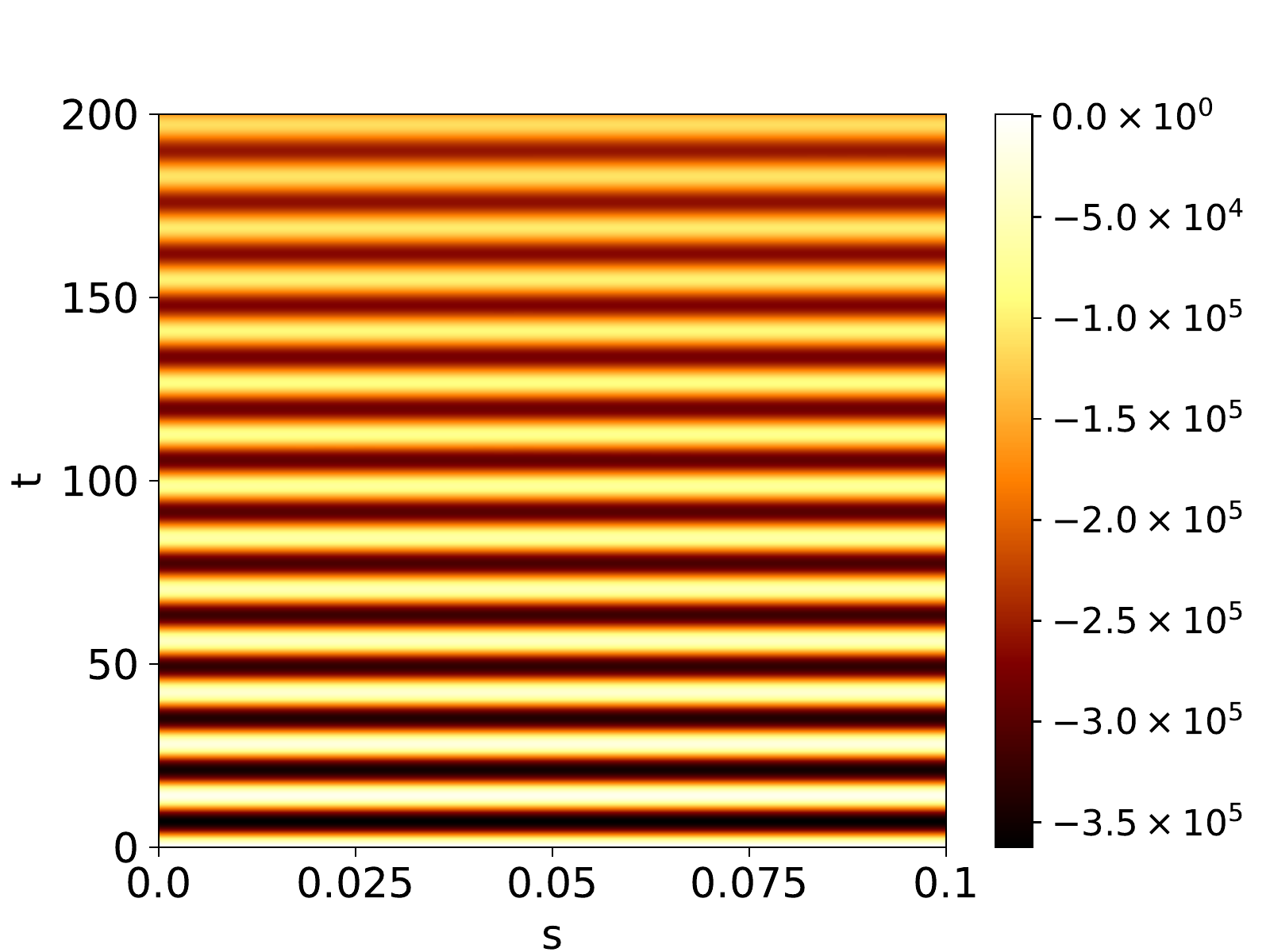}}\newline
    \subfloat{\includegraphics[width=.48\textwidth]{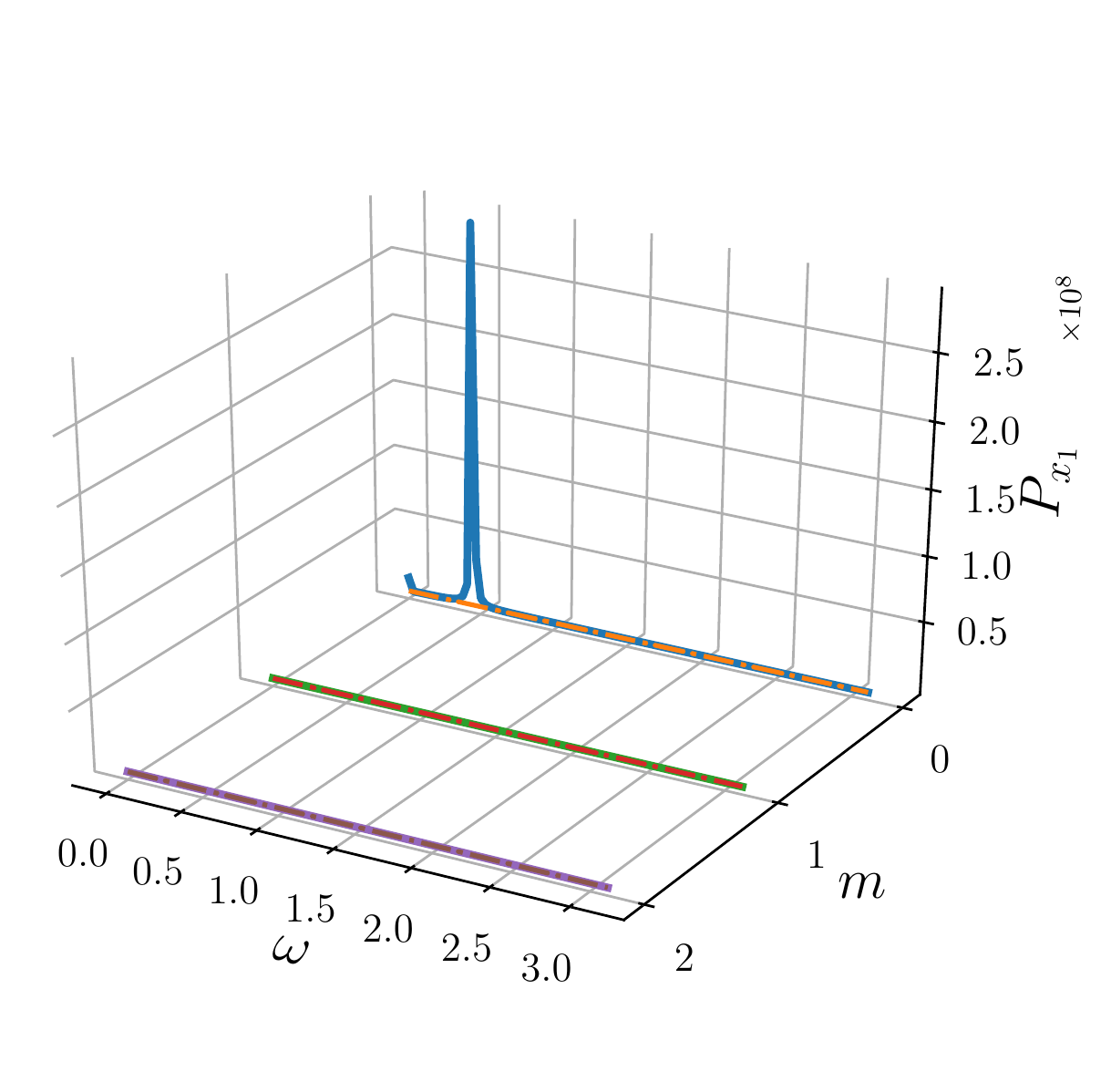}}
    \subfloat{\includegraphics[width=.48\textwidth]{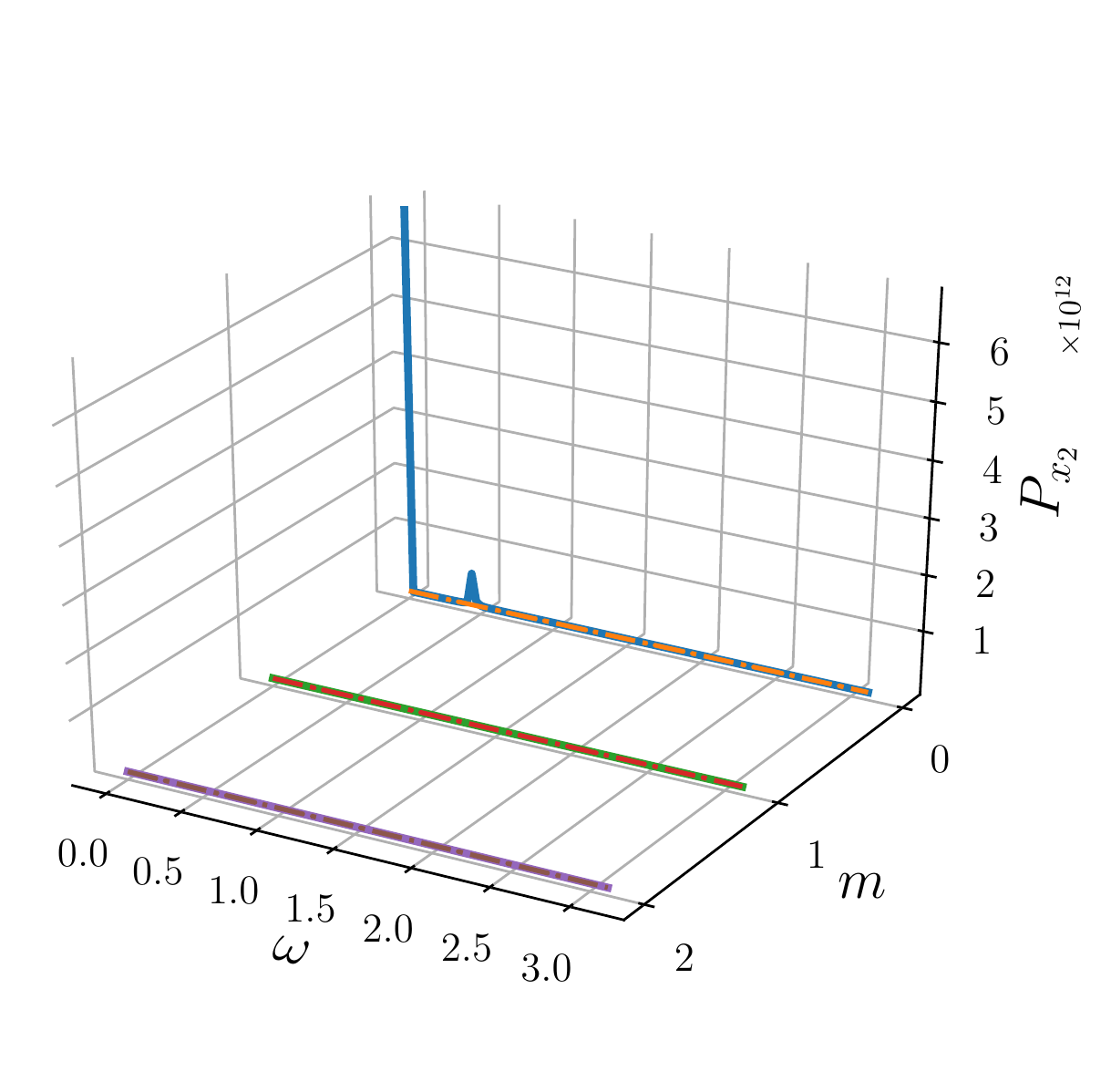}}
  \caption{The patterns and power spectra for $x_1$ (left) and $x_2$ (right) with noise generated by a violet noise process. The discrepancy in the power spectra originates from the truncation of the Gaussian process. The negative species concentrations indicate that a non-linear theory needs to be used to fully model a violet noise system. The spectra were averaged over 50 repetitions with simulation time $T=200$. The subsystem size was $\Omega = 5000$.}
  \label{Violet}
\end{figure}

To illustrate the behaviour of the system on the other end of the colour spectrum we simulated violet noise which stabilised the temporal oscillations further and due to large power at large $\omega$ drove the oscillation amplitude in the linear treatment to non biologically viable values. Again, this indicates that violet noise cannot be treated in biological applications with a simple linear theory, but a full non-linear theory must be used. For violet noise the effect of the truncation of the Gaussian became too large to get an accurate prediction of the power spectra as can be seen in Figure \ref{Violet}. Further, while noise colours with positive exponent are actively studied in biology \cite{Szendro2001}, the potential emergence of blue or violet noise in biological systems is not clear.

\subsection{Stochastic Auxiliary Networks}\label{OscSection}

\begin{figure}
  \centering
  \includegraphics[scale=0.5]{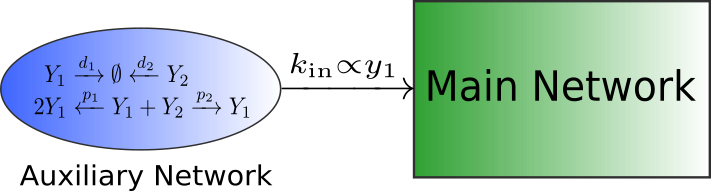}
  \caption{A schematic of an auxiliary network and its input to the main system. This is a specific example of scenario (b) of Figure \ref{Concepts}.}
  \label{AuxNetwork}
\end{figure}

We now proceed to the second major source of random inflows, namely the dependence of a stochastic auxiliary network. An auxiliary network is a chemical reaction network which provides an input into the Turing system (see Figure \ref{AuxNetwork}). The auxiliary network is connected to the main network only via an inflow reaction and, if the auxiliary network reaches a steady state, it can be subsumed into the zero complex for practical modelling. If, however, the auxiliary network exhibits more complex dynamics such as deterministic or stochastic oscillations, then it needs to be treated as a part of the main network. We focus on auxiliary networks which are deterministically stable, but show stochastic quasi-cycles. The dynamics of such systems are modelled by using a correlated stochastic inflow parameter as outlined in section \ref{RandomInflows}.

As an illustrative example we use the predator-prey system described in \cite{McKane2005},
\begin{align}
  Y_1\xrightarrow{d_1}&\emptyset\xleftarrow{d_2} Y_2,\nonumber\\
  2Y_1\xleftarrow{p_1} & Y_1+Y_2\xrightarrow{p_2} Y_1.
  \label{PredPreyEq}
\end{align}
In \cite{McKane2005} it was shown that in the deterministic regime the system \eqref{PredPreyEq} has exactly one attractor for any choice of rate constants. Hence, as the inflow parameters $k_1$ and $k_2$ are proportional to the concentration of $Y_1$, they will have a constant value. However, when the copy numbers of $Y_1$ and $Y_2$ are small and stochastic fluctuations are important the predator-prey model \eqref{PredPreyEq} can exhibit so-called ``quasi-cycles'' which are stochastic analogue of limit-cycles and manifest themselves as peaks in the power spectra.

Consider this ``predator-prey noise'' in a Schnakenberg system. Suppose the inflows $k_1$ and $k_2$ in the Schnakenberg system depend on the presence of the chemical species $Y_1$, such that $k_1 \propto y_1$ and $k_2\propto y_1$, where $y_1$ denotes the concentration of $Y_1$.
Again, we assume $g_{ij}(t,t') = g(t,t')$ for mathematical simplification. In Fourier space the power spectrum of $Y_1$ is
\begin{equation}
  P_\text{Predator-Prey} = \frac{\alpha + \beta\omega^2}{\left(\omega^2+\Omega^2_0\right)+\Gamma^2\omega^2},
  \label{PredPreySpecEq}
\end{equation}
where we use the parameter values from \cite{McKane2005} $\alpha = 0.000384,\: \beta = \Gamma = 0.04,\: \Omega_0^2 = 0.016$.

\begin{figure}
  \centering
    \subfloat{\includegraphics[width=.48\textwidth]{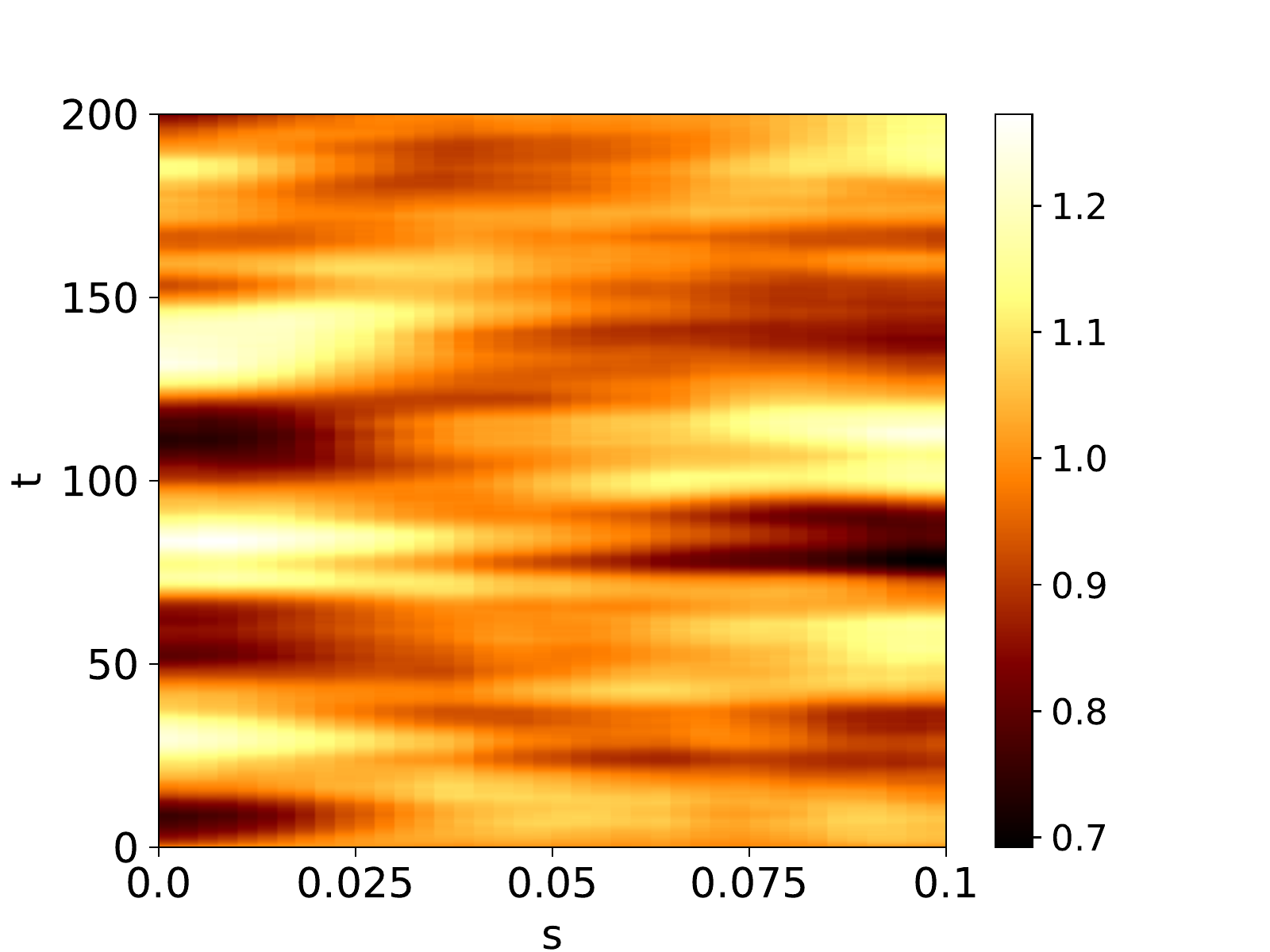}}
    \subfloat{\includegraphics[width=.48\textwidth]{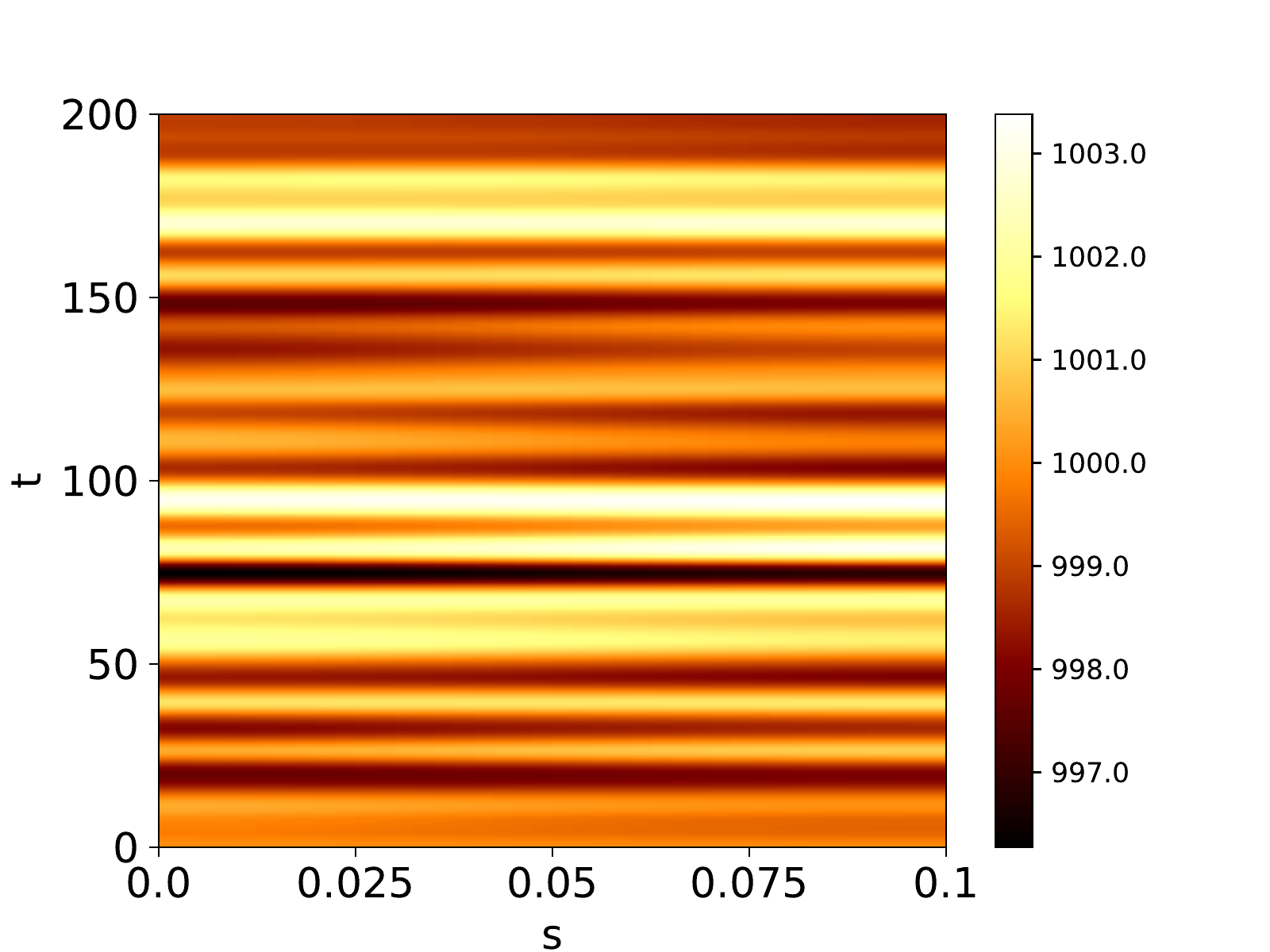}}\newline
    \subfloat{\includegraphics[width=.48\textwidth]{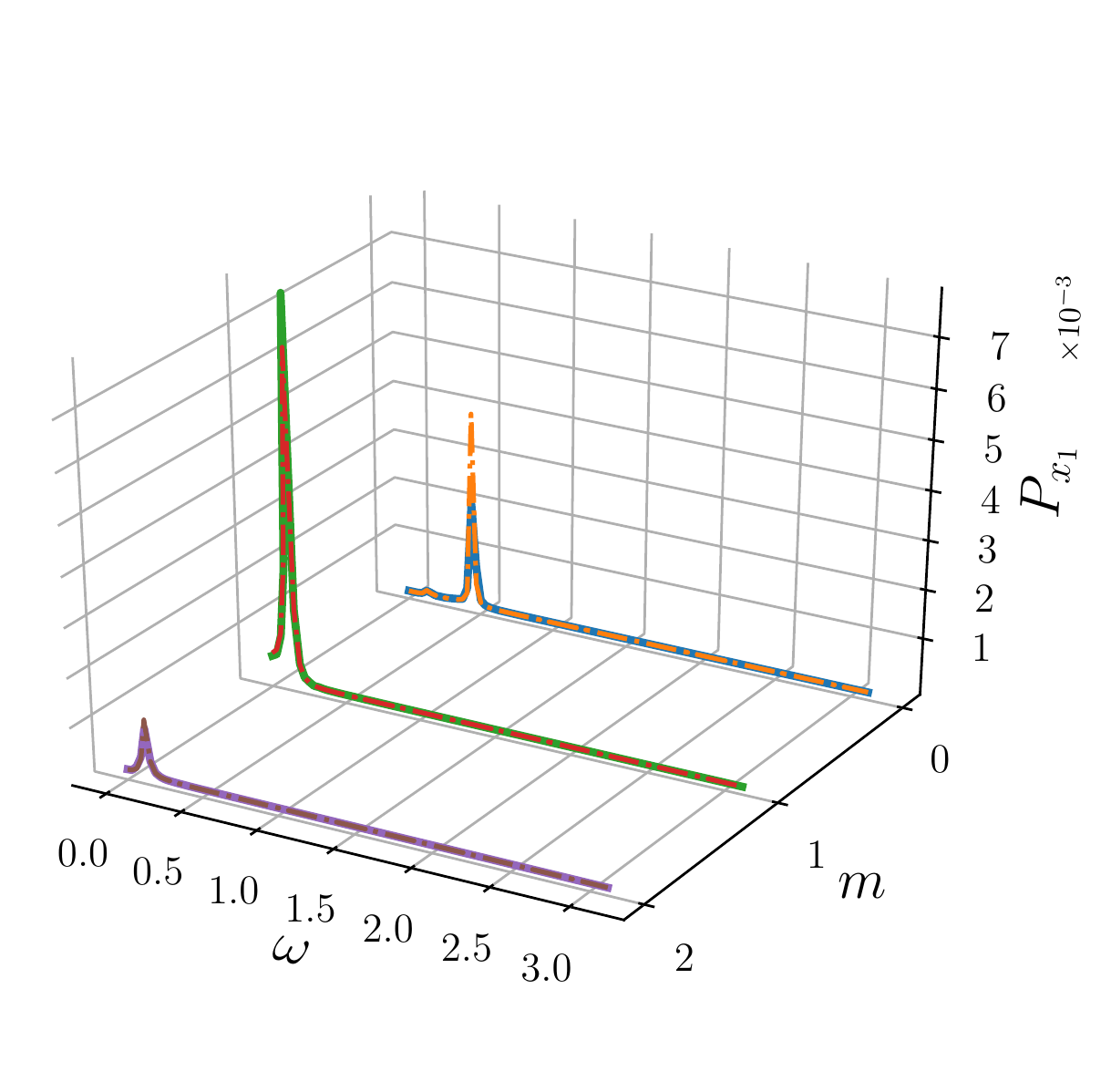}}
    \subfloat{\includegraphics[width=.48\textwidth]{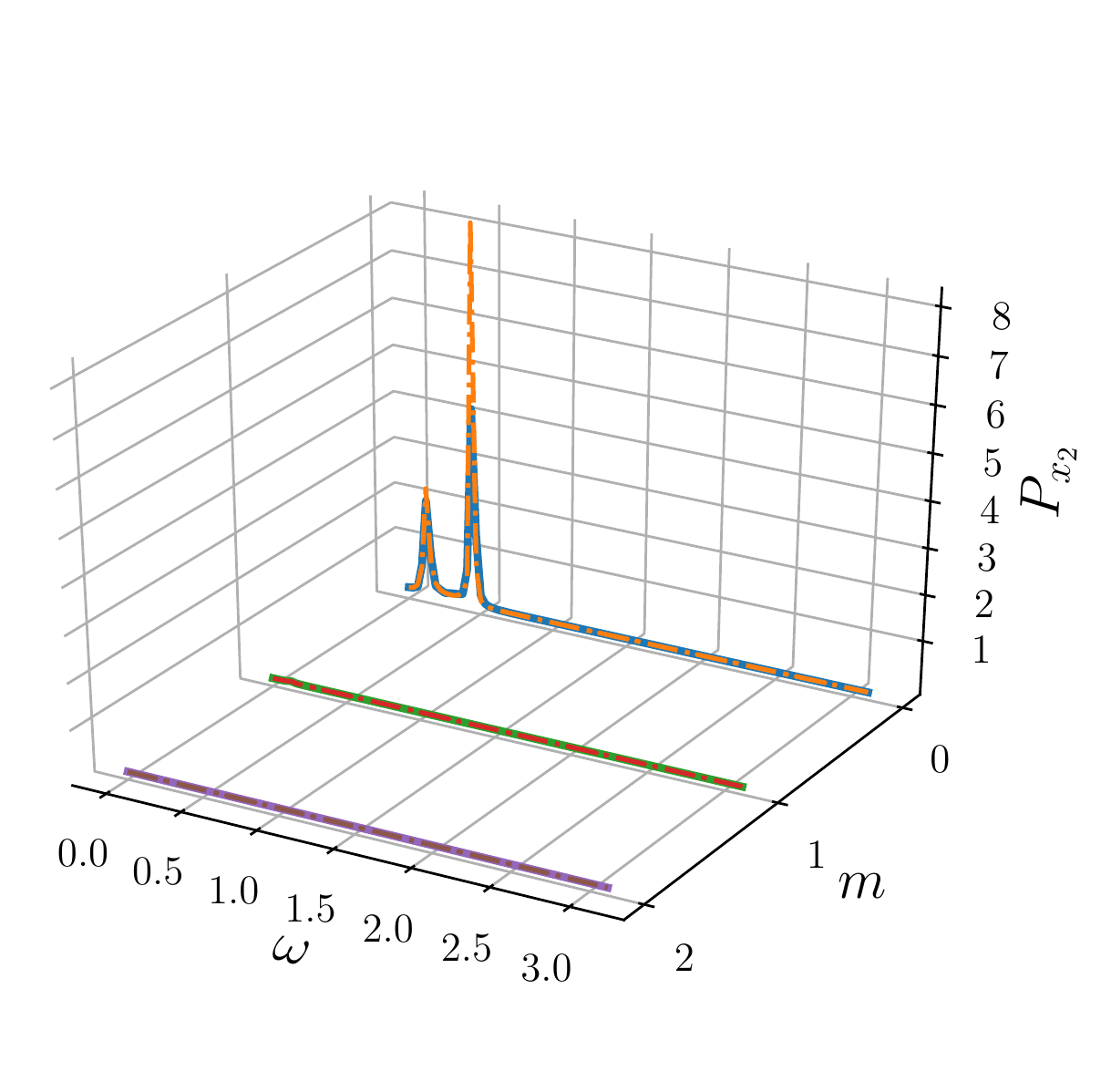}}
  \caption{The power spectra for the species $x_1$ (left) and $x_2$ (right) with noise generated by a stochastic predator-prey network \cite{McKane2005}. The power spectra inherited a second peak from the subsystem which activates oscillatory modes on top of the deterministic oscillation frequency. The spectra were averaged over 50 repetitions and simulation time $T=200$ with $m$ indexing the spatial mode. The subsystem size was $\Omega = 100$.}
  \label{PredPrey}
\end{figure}

Equation \eqref{PredPreySpecEq} has a peak at $\omega \sim 0.06$, so we expect peaks at a similar frequency for each non-zero spatial mode in the power spectrum of the Schnakenberg system.
The resulting patterns and the corresponding average power spectra are presented in Figure \ref{PredPrey}. The power spectra gained a second peak in the $m=0$ mode and peak in all modes $m>0$ at $\omega \sim 0.06$. Therefore, it can be seen that a deterministic parent system can inherit the dynamics of a stochastic auxiliary network and mix it with its own intrinsic dynamics.

\subsection{Mixed Noise}

Next, we investigated the case of mixing noise processes, in particular, we consider an Ornstein-Uhlenbeck process which has a different correlation time for $x_1$ and $x_2$, $\tau_1$ and $\tau_2$ respectively. Then we let $\tau_2 \to 0$ in order to recover the white noise case for species two.

We introduce the auxiliary normal stochastic processes $\xi$ and hence, the spatial modes of the  noise $\eta_\kappa$ are described by \cite{Hanggi1995}
\begin{equation}
  \frac{d\eta_\kappa}{dt} = -\begin{pmatrix}
  \tfrac{1}{\tau_1}& 0\\
  0 & \tfrac{1}{\tau_2}
\end{pmatrix}\eta_\kappa + \begin{pmatrix}
  \tfrac{1}{\tau_1}& 0\\
  0 & \tfrac{1}{\tau_2}
  \end{pmatrix}b_\kappa\xi,
  \label{OrUlAuxEq}
\end{equation}
where $b_\kappa$ is a $2 \times 4$ matrix satisfying $b_\kappa b_\kappa^T = B_\kappa$. We Fourier transform \eqref{OrUlAuxEq} and define the matrix,
\begin{equation}
  \phi = \begin{pmatrix}
  \tfrac{1}{\tau_1}+i\omega & 0\\
  0 & \tfrac{1}{\tau_2}+i\omega
  \end{pmatrix}
\end{equation}
to give
\begin{equation}
  \eta_\kappa = \phi^{-1}\begin{pmatrix}
    \tfrac{1}{\tau_1}& 0\\
    0 & \tfrac{1}{\tau_2}
    \end{pmatrix}b\xi.
\end{equation}
Letting $\tau_2\to 0$ and computing the covariance matrix gives
\begin{equation}
  N = \langle\eta_\kappa(\omega)\eta_\kappa(\omega)^\dagger\rangle = \begin{pmatrix}
  B_{k,11}\tfrac{1}{1+\omega^2\tau_1^2} & B_{k,12}\tfrac{1}{1+i\omega\tau_1}\\
  B_{k,12}\tfrac{1}{1-i\omega\tau_1} & B_{k,22}
  \end{pmatrix},
\end{equation}
where $B_{k,ij}$ are the $ij^\text{th}$ elements of the $B_\kappa$ matrix. Note that $N$ is hermitian and therefore we expect real power spectra for the patterns of $x_1$ and $x_2$.

\begin{figure}
 \centering
    \subfloat{\includegraphics[width=.48\textwidth]{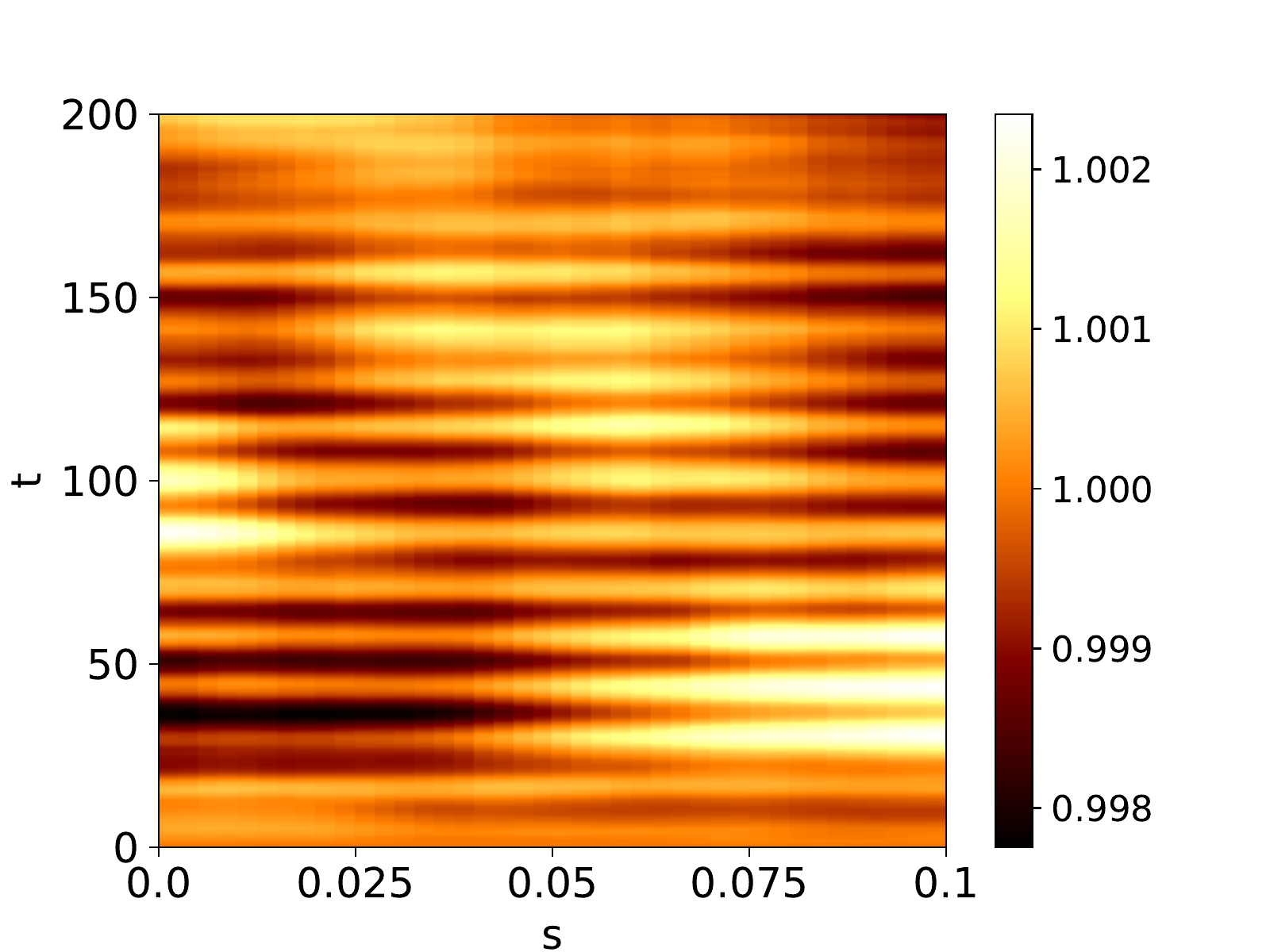}}
    \subfloat{\includegraphics[width=.48\textwidth]{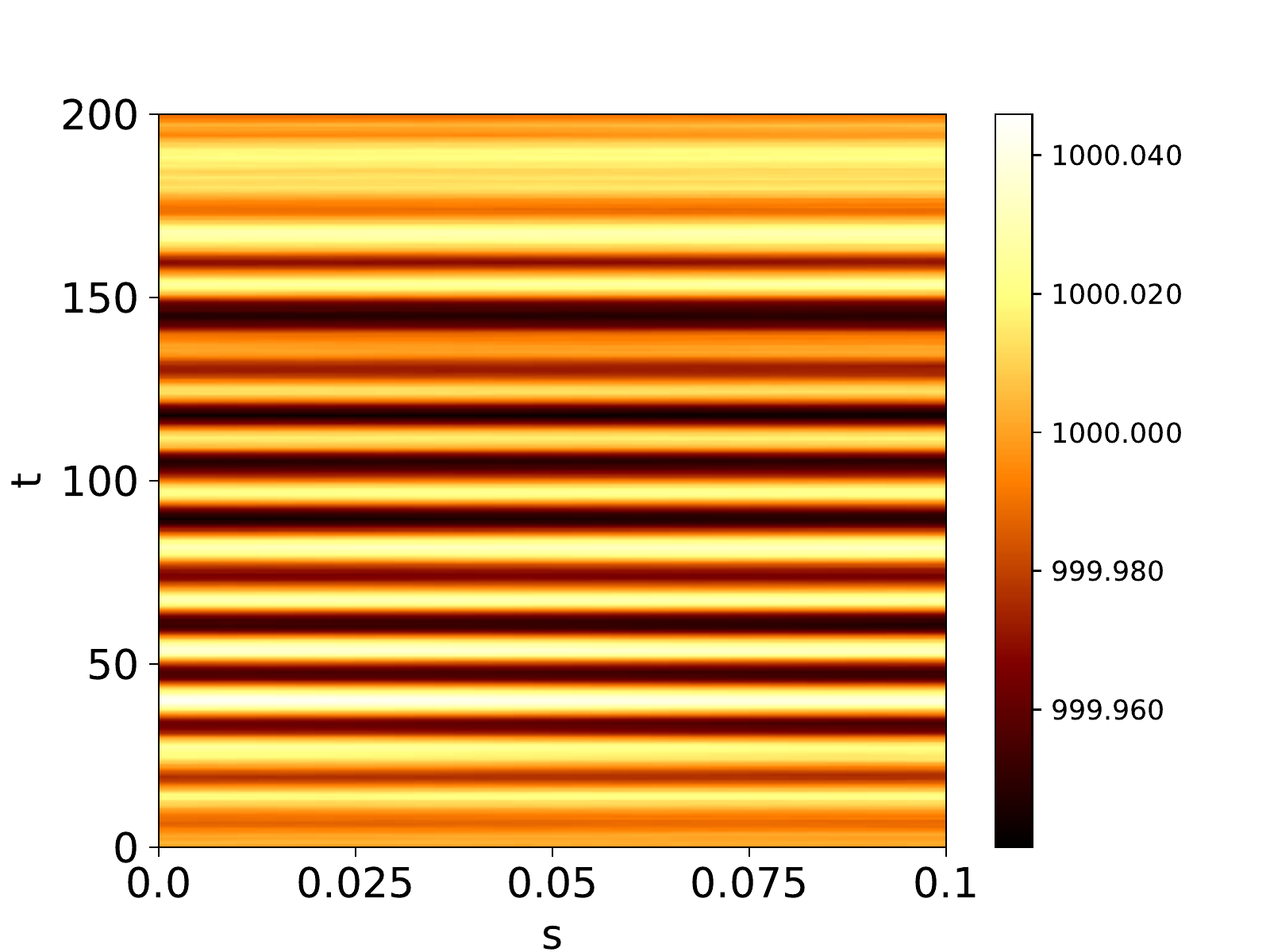}}\newline
    \subfloat{\includegraphics[width=.48\textwidth]{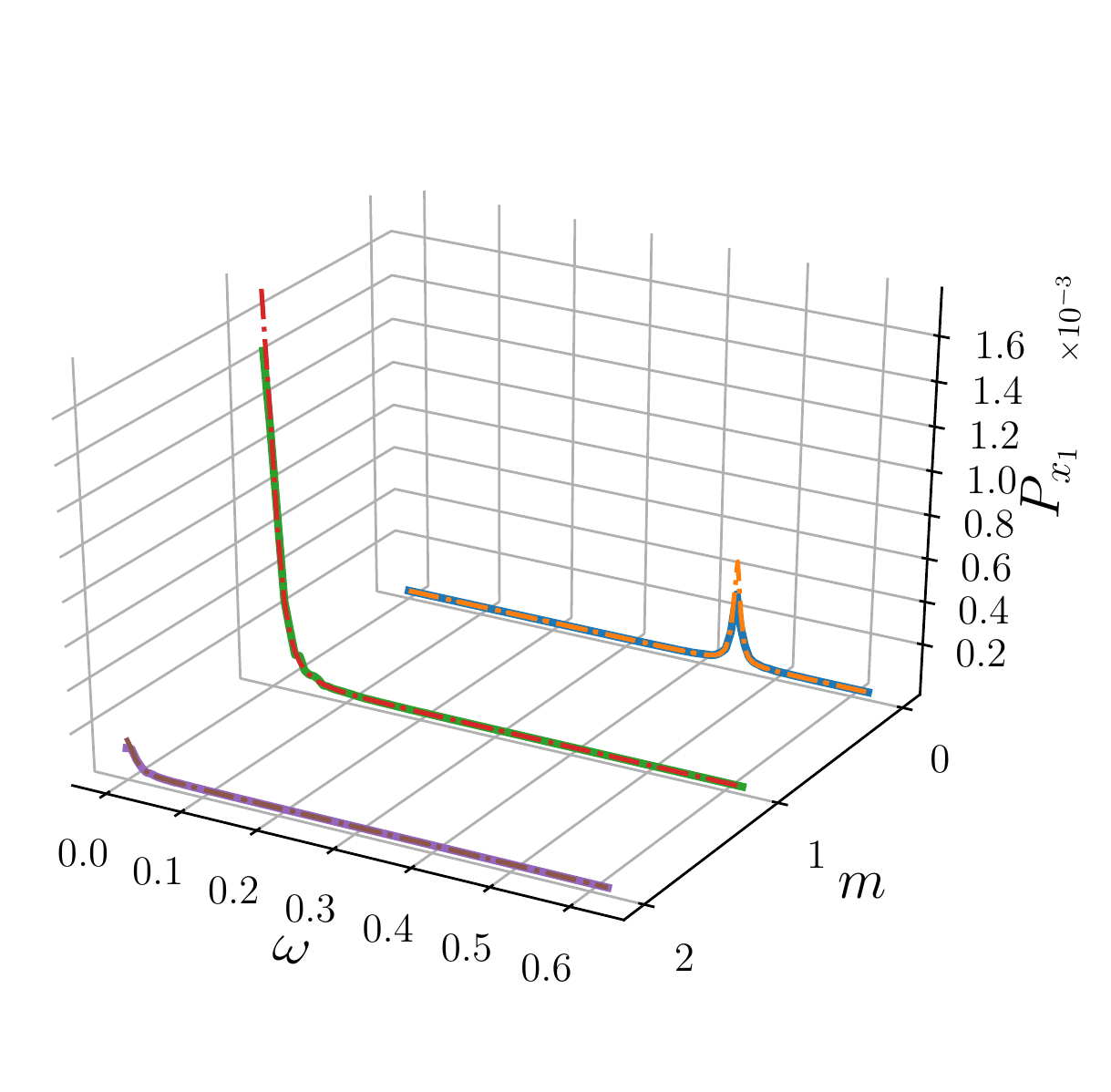}}
    \subfloat{\includegraphics[width=.48\textwidth]{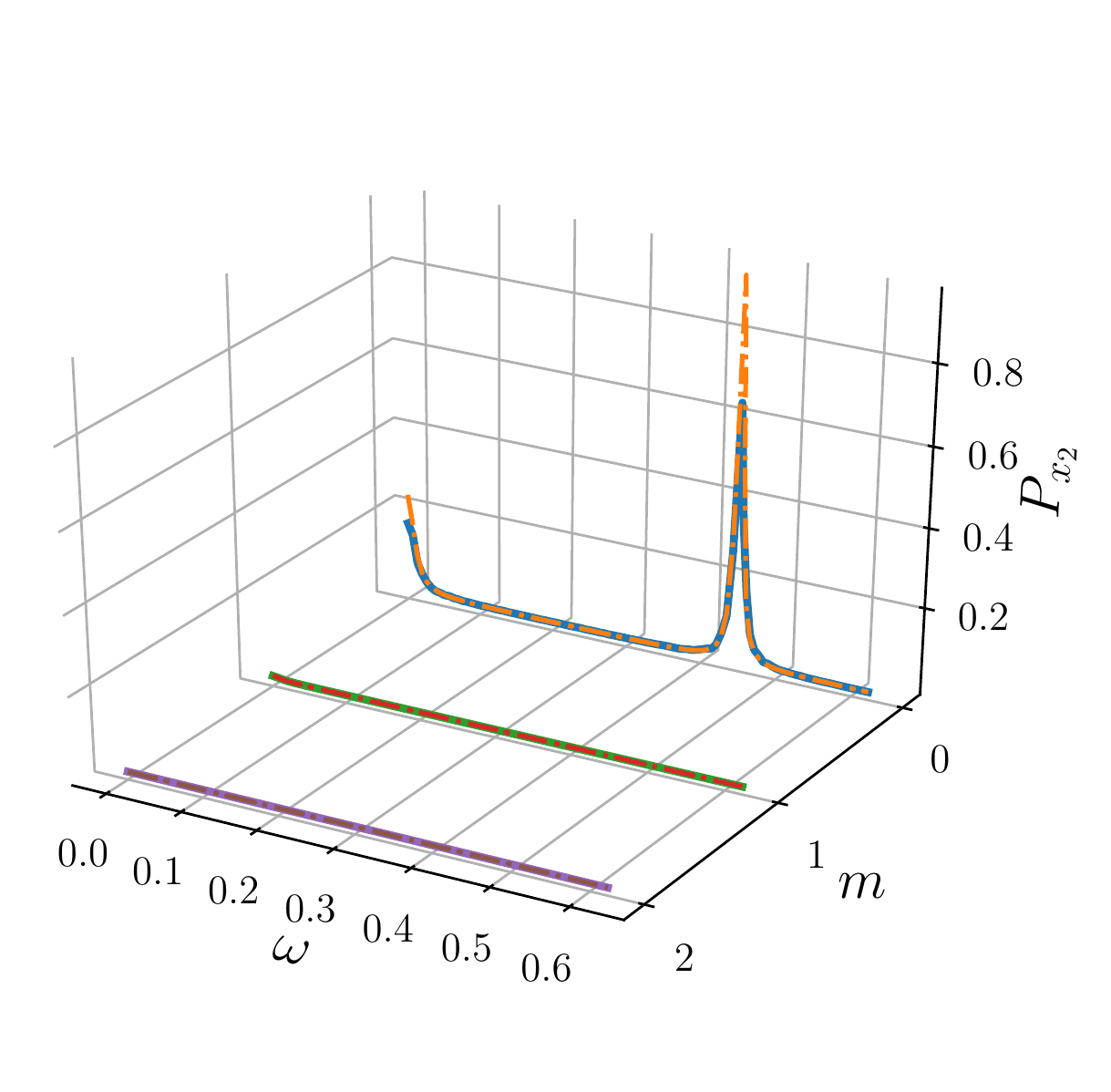}}
  \caption{The patterns generated for the species $x_1$ (left) and $x_2$ (right) with Ornstein-Uhlenbeck noise and $\tau = 100$ for $x_1$ and $x_2$ with white noise.The patterns are phenomenologically similar to the Ornstein-Uhlenbeck patterns, however, with reduced amplitude. The power spectrum of $x_2$ is a mixture of White noise and Ornstein-Uhlenbeck noise. The power spectra were averaged over 50 simulations with simulation time $T=1000$ and subsystem size $\Omega=5000$.}
  \label{Mixed}
\end{figure}

Simulating equation \eqref{StochNetLinFinal} with the noise process described by \eqref{OrUlAuxEq} gives rise to the mixed patterns seen in Figure \ref{Mixed}. Note, that computationally the limit $\tau_2\to 0$ is equal to setting $\tau_2$ to the time step $dt$. The patterns of both species appear similar to the pure Ornstein-Uhlenbeck patterns, however, with reduced amplitude.

\subsection{Reduced Stochastic Inflows}

In the final subsection we instigate the case of only species $x_1$ being subjected to stochastic inflows. The species $x_2$ is assumed to only have deterministic inflow at a rate $k_2$ and $\eta_2 = 0$. Hence, the correlation matrices $N$ will be reduced to
\begin{equation}
  N = \begin{pmatrix}
  \langle\eta_{1k}\eta_{1k}^*\rangle & 0\\
  0 & 0
  \end{pmatrix}.
\end{equation}

In the noise processes studied in this paper the behaviour of the species is virtually unchanged and the patterning is robust with respect to disregarding cross-correlations.

\section{Conclusion}\label{Conclusion}

In this paper we investigated the effect of stochastic inflows on a deterministic reaction-diffusion system. We restricted the class of networks considered to monostationary systems and used results from real algebraic geometry to show how monostationarity is related to network structure. We then introduced a stochastic perturbation to an inflow reaction as a truncated Gaussian process with the expectation value at the deterministic inflow parameter. After linearising we computed the power spectra for arbitrary noise colours and showed that in simple cases the power spectra can be derived as a multiplicative factor. The remainder of the paper consisted of applying our analysis to the Schnakenberg system and highlighting the effects various noise colouring can have on a reaction-diffusion system.

We briefly restated the results from the white noise analysis, proceeding to add coloured noise and then demonstrating its effect by computing the power spectra. In particular, for the simple case when all species experience the same temporal correlations, the power spectra of the correlations appear as a multiplicative factor in the total power spectra. Hence, depending on the nature of the correlations they will suppress or excite temporal frequencies at all spatial modes.  Ornstein-Uhlenbeck noise has a Lorentzian frequency distribution and, therefore, suppresses positive frequencies. Power law noise can either completely stabilise or destabilise the pattern depending on the sign of the exponent, $\alpha$. This is due to the fact that for positive $\alpha$ temporal frequencies at $\omega = 0$ go to infinity and, therefore, all oscillations are suppressed which results in a stable pattern which resembles a pattern arising from simulations inside the Turing regime.

For the simulation times used in this paper, the noise vector was actually constant, and, therefore, deterministic methods could be used to study the stabilising effect of pink or red noise. The opposite is the case for negative $\alpha$ where the frequency contribution in the power spectrum increases as $\omega$ increases and oscillations are enhanced. However, certain aspects of the red, blue, or violet, noise cases cannot be explained by our simple analytic prediction. The biological significance and the correct analytic description of blue/violet noise could form a part of further work.

When turning to stochastic auxiliary networks we see that the reaction-diffusion system can inherit traits of the dynamics of the auxiliary network. In particular, we showed that when the auxiliary network exhibits a predator-prey dynamic with stochastic oscillations these oscillations can still be observed in the deterministic main network. We concluded the range of applications by considering mixed noise with Ornstein-Uhlenbeck input for species $x_1$ and white noise for species $x_2$. In this case, the system behaves similar to a system with pure Ornstein-Uhlenbeck dynamics and differences can only be found by a power spectral analysis. Finally, we observed in the cases considered that when only one species experiences stochastic inflow the patterns created are, except for potential special cases, similar to the ones when the inflow to both species is randomised. However, the extent to which such results may hold in generality is for further work.

Further directions could also include attempting to relate these studies to potential mechanisms of left-right symmetry breaking amplification in developmental biology, in particular the impact of induced Nodal production on one side of the embryonic node, which is hypothesised to be driven by ciliary fluid flows and also highly error-prone \cite{Blum2014}. It is generally asserted that the resulting interactions of the gene products Nodal and Lefty, which are major contenders as Turing morphogens \cite{Muller2012,Chen2002,Schier2003}, amplify this initial error-prone signal to generate robust patterning, driving downstream developmental left-right asymmetry. However, the ability of Turing systems to amplify the spatially localised, error-prone, and thus stochastic, influx of activator morphogen \cite[Figure 2]{Blum2014} to induce a robust symmetry breaking self-organisation, as well as any additional constraints required to do so, is theoretically untested. Thus examining the mechanistic basis of these postulates in this critical developmental biological process provides a fundamental application for the theoretical foundations developed here.

\bibliographystyle{unsrt}
\bibliography{Bibliography}

\begin{thebibliography}{10}

\bibitem{Murray2008}
J.~D. Murray.
\newblock {\em {Mathematical Biology II - Spatial Models and Biomedical
  Applications}}, volume~18 of {\em Interdisciplinary Applied Mathematics}.
\newblock Springer-Verlag, New York, 3 edition, 2008.

\bibitem{Hochberg2003}
D.~Hochberg, F.~Lesmes, F.~Mor{\'{a}}n, and J.~P{\'{e}}rez-Mercader.
\newblock {\em Phys. Rev. E}, 68(6), 2003.

\bibitem{Turing37}
A.~Turing.
\newblock {\em Philos. Trans. R. Soc. London B Biol. Sci.}, 237(641):37--72,
  1952.

\bibitem{McKane2014}
A.~J. McKane, T.~Biancalani, and T.~Rogers.
\newblock {\em Bull. Math. Biol.}, 76(4):895--921, 2014.

\bibitem{Butler2011}
T.~Butler and N.~Goldenfeld.
\newblock {\em Phys. Rev. E}, 84:011112, 2011.

\bibitem{Xavier2013}
D.~Xavier.
\newblock {\em {On the theory of cell migration: durotaxis and chemotaxis}}.
\newblock PhD thesis, Universitat Politcnica de Catalunya, 2013.

\bibitem{Schumacher2013}
L.~J. Schumacher, T.~E. Woolley, and R.~E. Baker.
\newblock {\em Phys. Rev. E}, 87(4):042719, 2013.

\bibitem{McKane2005}
A.~J. McKane and T.~J. Newman.
\newblock {\em Phys. Rev. Lett.}, 94(21):218102, 2005.

\bibitem{Biancalani2011}
T.~Biancalani, T.~Galla, and A.~J. McKane.
\newblock {\em Phys. Rev. E}, 84:026201, 2011.

\bibitem{woolley2012effects}
T.~E. Woolley, R.~E. Baker, E.~A. Gaffney, P.~K Maini, and S.~Seirin-Lee.
\newblock {\em Phys. Rev. E}, 85(5):051914, 2012.

\bibitem{woolley2011influence}
T.~E. Woolley, R.~E. Baker, E.~A. Gaffney, and P.~K. Maini.
\newblock {\em Phys. Rev. E}, 84(4):041905, 2011.

\bibitem{Gillespie2000}
D.~T. Gillespie.
\newblock {\em Jour. Chem. Phys.}, 113(1):297--306, 2000.

\bibitem{Dauxois2009}
T.~Dauxois, F.~{Di Patti}, D.~Fanelli, and A.~J. McKane.
\newblock {\em Phys. Rev. E}, 79(3):036112, 2009.

\bibitem{Horn1972}
F.~Horn and R.~Jackson.
\newblock {\em Arch. Rat. Mech. and Anal.}, 47(2):81--116, 1972.

\bibitem{woolley2017turing}
T.~E. Woolley, R.~E. Baker, and P.~K. Maini.
\newblock In {\em The Incomputable}, pages 219--235. Springer, 2017.

\bibitem{Muller2016}
S.~M{\"{u}}ller, E.~Feliu, G.~Regensburger, C.~Conradi, A.~Shiu, and
  A.~Dickenstein.
\newblock {\em Found. Comput. Math.}, 16(1):69--97, 2016.

\bibitem{Castets1990}
V.~Castets, E.~Dulos, J.~Boissonade, and P.~De~Kepper.
\newblock {\em Phys. Rev. Lett.}, 64:2953--2956, 1990.

\bibitem{Cartwright2002}
J.~H.~E. Cartwright.
\newblock {\em eprint arXiv:nlin/0211001}, 2002.

\bibitem{Kondo2010}
Sh. Kondo and T.~Miura.
\newblock {\em Science}, 329(5999):1616--1620, 2010.

\bibitem{Picco2017}
N.~Picco, E.~Sahai, P.~K. Maini, and A.~R.~A. Anderson.
\newblock {\em Cancer Research}, 77(19):5409--5418, 2017.

\bibitem{Lenive2016}
O.~Lenive, P.~D.~W. Kirk, and M.~P.~H. Stumpf.
\newblock {\em BMC Syst. Biol.}, 10(1):81, aug 2016.

\bibitem{Feinberg1987}
M.~Feinberg.
\newblock {\em Chem. Eng. Sci.}, 42(10):2229--2268, 1987.

\bibitem{Feinberg1988}
M.~Feinberg.
\newblock {\em Chem. Eng. Sci.}, 43(1):1--25, 1988.

\bibitem{Banerjee2014}
K.~{Banerjee} and K.~{Bhattacharyya}.
\newblock {\em ArXiv e-prints}, 2014.

\bibitem{Joshi2014}
B.~Joshi and A.~Shiu.
\newblock {\em ArXiv e-prints}, 2014.

\bibitem{Felix2016}
B.~F{\'{e}}lix, A.~Shiu, and Z.~Woodstock.
\newblock {\em Appl. Math. Comput.}, 287-288:60--73, 2016.

\bibitem{Banaji2018}
M.~Banaji and C.~Pantea.
\newblock {\em SIAM J. Appl. Math.}, 78(2):1105--1130, 2018.

\bibitem{Frank2013}
S.~A. Frank.
\newblock {\em Biol. Direct}, 8:31, 2013.

\bibitem{Kampen2007}
N.~G. van Kampen.
\newblock {\em {Stochastic Processes in Physics and Chemistry}}.
\newblock Elsevier, 1983.

\bibitem{Adamer2017}
M.~F. Adamer, T.~E. Woolley, and H.~A. Harrington.
\newblock {\em Jour. Roy. Soc. Interface}, 14(137), 2017.

\bibitem{Woolley2011a}
T.~E. Woolley, R.~E. Baker, E.~A. Gaffney, and P.~K. Maini.
\newblock {\em Phys. Rev. E}, 84(4):046216, 2011.

\bibitem{Smith1985}
G.~D. Smith.
\newblock {\em {Numerical solution of partial differential equations, Finite
  difference methods}}.
\newblock Clarendon Press, 1985.

\bibitem{Briggs1995}
W.~Briggs and V.~Henson.
\newblock {\em {The DFT: An Owner's Manual for the Discrete Fourier
  Transform}}.
\newblock Society for Industrial and Applied Mathematics, 1995.

\bibitem{Woolley2011b}
T.~E. Woolley, R.~E. Baker, E.~A. Gaffney, and P.~K. Maini.
\newblock {\em Phys. Rev. E}, 84:021915, 2011.

\bibitem{Iron2004}
D.~Iron, J.~Wei, and M.~Winter.
\newblock {\em J. Math. Biol.}, 49(4):358--390, 2004.

\bibitem{Maini2012}
P.~K. Maini, T.~E. Woolley, R.~E. Baker, E.~A. Gaffney, and S.~S. Lee.
\newblock {\em J. Roy. Soc. Interface Focus}, 2(4):487--496, 2012.

\bibitem{Ward2002}
M.~J. Ward and J.~Wei.
\newblock {\em Stud. Appl. Math.}, 109(3):229--264, 2002.

\bibitem{Flach2007}
E.~H. Flach, S.~Schnell, and J.~Norbury.
\newblock {\em Appl. Math. Lett.}, 20(9):959--963, 2007.

\bibitem{Kloeden1995}
Peter~E. Kloeden and Eckhard. Platen.
\newblock {\em {Numerical solution of stochastic differential equations}}.
\newblock Springer-Verlag, 1995.

\bibitem{Milshtein1994}
G~N Milshtein and M~V {Tret 'yakov}.
\newblock {Numerical Solution of Differential Equations with Colored Noise}.
\newblock {\em J. Stat. Phys.}, 774(3), 1994.

\bibitem{Timmer1995}
J.~{Timmer} and M.~{Koenig}.
\newblock {\em Astronomay and Astrophysics}, 300:707, August 1995.

\bibitem{Hanggi1995}
P.~Hanggi and P.~Jung.
\newblock {\em Adv. Chem. Physics. Vol LXXXIX}, LXXXIX:239--326, 1995.

\bibitem{Cates2015}
M.~E. Cates and J.~Tailleur.
\newblock {\em Annu. Rev. Condens. Matter Phys.}, 6(1):219--244, 2015.

\bibitem{Filippi2015}
S.~Filippi, C.~Barnes, P.~Kirk, T.~Kudo, S.~McMahon, T.~Tsuchiya, T.~Wada, Sh.
  Kuroda, and M.~Stumpf.
\newblock {\em bioRxiv},
  https://www.biorxiv.org/content/early/2015/07/01/021790, 2015.

\bibitem{Szendro2001}
P.~Szendro, G.~Vincze, and A.~Szasz.
\newblock {\em Electro- and Magnetobiology}, 20(2):215--229, 2001.

\bibitem{Blum2014}
M.~Blum, K.~Feistel, T.~Thumberger, and A.~Schweickert.
\newblock {\em Development}, 141(8):1603--1613, 2014.

\bibitem{Muller2012}
P.~M{\"u}ller, K.~W. Rogers, B.~M. Jordan, J.~S. Lee, D.~Robson, S.~Ramanathan,
  and A.~F. Schier.
\newblock {\em Science}, 336(6082):721--724, 2012.

\bibitem{Chen2002}
Y.~Chen and A.~F. Schier.
\newblock {\em Curr. Biol.}, 12(24):2124--2128, 2002.

\bibitem{Schier2003}
A.~F. Schier.
\newblock {\em Annu. Rev. Cell Dev. Biol.}, 19(1):589--621, 2003.

\end{thebibliography}
\end{document}